\documentclass[11pt]{article}

\usepackage{graphicx}
\usepackage{amsmath}
\usepackage{xcolor}
\usepackage{ragged2e}
\usepackage{amssymb}
\usepackage{tabularx}
\usepackage{booktabs}
\usepackage{subfig}
\usepackage{empheq}
\usepackage{stackrel}
\usepackage[font={small,it}]{caption}
\usepackage{hyperref}
\usepackage{url}
\usepackage{textcomp}
\usepackage{upgreek}

\newcommand{\myquote}[1]{``#1''}
\newcommand{\leftrarrows}{\mathrel{\raise.75ex\hbox{\oalign{%
  $\scriptstyle\leftarrow$\cr
  \vrule width0pt height.5ex$\hfil\scriptstyle\relbar$\cr}}}}
\newcommand{\lrightarrows}{\mathrel{\raise.75ex\hbox{\oalign{%
  $\scriptstyle\relbar$\hfil\cr
  $\scriptstyle\vrule width0pt height.5ex\smash\rightarrow$\cr}}}}
\newcommand{\Rrelbar}{\mathrel{\raise.75ex\hbox{\oalign{%
  $\scriptstyle\relbar$\cr
  \vrule width0pt height.5ex$\scriptstyle\relbar$}}}}

\definecolor{myred}{rgb}{0.82, 0.1, 0.26}
\definecolor{mygrey}{rgb}{0.6, 0.6, 0.6}
\newcommand{\rev}[1]{{\color{black}#1}}
\newcommand{\gbrev}[1]{{\color{black}#1}}

\newcommand{\entry}[2]{\begin{tabular}{p{4.75em} l} #1 & #2\end{tabular}\\}

\title{\bf Core-shell enhanced single particle model for lithium iron phosphate batteries: model formulation and analysis of numerical solutions}
\author{Gabriele Pozzato$^*$,  Aki Takahashi\thanks{Energy Resources Engineering, Stanford University, Stanford, CA 94305. } ,  Xueyan Li\thanks{Assistant manager, Research Team, Tech Center,  LG Energy Solution Michigan, Troy,  MI 48083} ,  Donghoon Lee\thanks{Specialist, Data Modeling Algorithm Team, Battery R{\&}D,  LG Energy Solution, Daejeon R{\&}D Campus,  South Korea 34122} , \\ Johan Ko\thanks{Professional, Data Modeling Algorithm Team, Battery R{\&}D,  LG Energy Solution, Gwacheon R{\&}D Campus,   South Korea 13818} , and  Simona Onori$^{*,}$\thanks{Corresponding author,  {\tt\small sonori@stanford.edu}}} \date{}

\makeatletter
\def\leftrightarrowsfill@{\arrowfill@\leftrarrows\Rrelbar\lrightarrows}
\newcommand{\xleftrightarrows}[2][]{\ext@arrow 3399\leftrightarrowsfill@{#1}{#2}}
\makeatother

\begin{document}
\maketitle	

\begin{abstract}
\noindent
In this paper,  a core-shell enhanced single particle model for iron-phosphate battery cells is formulated,  implemented,  and verified.  Starting from the description of the positive and negative electrodes charge and mass transport dynamics, the positive electrode intercalation and deintercalation phenomena and associated phase transitions are described with the core-shell modeling paradigm.  Assuming two phases are formed in the positive electrode,  one rich and one poor in lithium,  a core-shrinking problem is formulated and the phase transition is modeled through a shell phase that covers the core one.  A careful discretization of the coupled partial differential equations is proposed and used to convert the model into a system of ordinary differential equations.  To ensure robust and accurate numerical solutions of the governing equations,  a sensitivity analysis of numerical solutions is performed and the best setting,  in terms of solver tolerances,  \rev{solid phase concentration discretization points},  and input current sampling time, is determined in a newly developed probabilistic framework.  Finally,  unknown model parameters are identified at different C-rate scenarios and the model is verified against experimental data.
\end{abstract}

\section{Introduction}
\rev{Electrification and the transition to clean and green energy  and circular economy have found their solution in lithium-ion battery (LIB) technology.   Lithium iron phosphate batteries (or LFP),  among the first LIBs to be commercialized \cite{padhi1997phospho}, are today standard in China, used mostly for electric scooters and small electric vehicles (EVs). LFP batteries use lithium iron phosphate ($\mathrm{LiFePO}_4$) as the cathode material alongside a graphite carbon electrode as the anode \cite{padhi1997phospho}.  LFP batteries do not decompose at higher temperatures, thus providing  thermal and chemical stability, which results in an intrinsically safer cathode material than other commercially available chemistries such as NMC and LCO batteries \cite{energygovblue}. 
The superior safety and charge/discharge rate characteristics of $\mathrm{LiFePO}_4$ have made it an attractive positive electrode material for LIB. In particular, the highly reversible transition between the olivine lithiated ($\mathrm{LiFePO}_4$) and delithiated ($\mathrm{FePO_4}$)  phases allows thousands of charge/discharge cycles with high rate capability at an ideal voltage plateau around 3.45 V vs. lithium metal.}

\rev{Specifically, during charge (and discharge) three phases are observed on the OCV  vs Amph-sec characteristic: 1)  Li-rich phase (only $\mathrm{LiFePO}_4$), 2) two-phase transition where the $\mathrm{LiFePO}_4$/$\mathrm{FePO_4}$ phases coexist, and 3) Li-poor phase (only $\mathrm{FePO}_4$), \cite{Yamada2005,delmas2011lithium,lim2016origin}.}

Developing a model incorporating the description of these two phases is crucial to understand lithium intercalation and deintercalation in LFP batteries and gain information on the electrochemical states from which one can build $SOC$ estimators.  
 
\rev{According to \cite{love2013}, conflicting models exist to describe the phase transition mechanisms, lithium insertion mechanisms, and the existence of a two-phase region.
In \cite{dreyer2010thermodynamic},  a many-particle model is proposed where  lithium is exchanged between individual particles, and sequential lithiation and delithiation are demonstrated. Transport and kinetic equations are ignored,  making the model unsuitable when a careful description of the electrochemical states is needed or at high C-rate.  The authors of \cite{delmas2008} propose a domino-cascade model to describe  lithium intercalation and deintercalation in the positive electrode lattice.  In this model,  the phase transition is modeled with a front moving, without any energy barrier,  inside the crystalline structure.  Mosaic models have also been investigated.  In this context,  smaller nucleation sites within a larger single particle,  each undergoing a phase change during charge and discharge, are considered \cite{andersson2001}.}

\rev{To model the $\mathrm{LiFePO_4/FePO_4}$ phase transitions,  in \cite{srinivasan2004discharge} a  core-shell approach,  similar to the one shown in \cite{subramanian2000shrinking} for the discharge of a metal hydride electrode,  is proposed. } At the  cost of adding some complexity to the electrochemical model -- i.e., a mass balance equation -- this technique allows for a detailed description of the electrode lithiation and delithiation dynamics.  Diffusion is assumed to be isotropic and phase transitions are described upon a shell and core phase interacting via a moving boundary.  To describe the path dependence during the battery operation,  the core-shell model prescribes different phases to the shell and core depending on whether the battery is charging or discharging. In \cite{koga2017state} and \cite{li2015modeling},  it is demonstrated the successful implementation of this core-shell model for the single particle model (SPM) and pseudo-two-dimensional (P2D),  respectively.

In this work, a core-shell enhanced single particle model (ESPM) is formulated to model two-phase transition dynamics during charge and discharge in $\mathrm{LiFePO_4}$/graphite batteries.  By means of the core-shell modeling principle, both the one-phase at the beginning (Li-rich)  and end (Li-poor) of charge (or discharge) and the two-phase region are modeled. 
A careful discretization of the electrochemical partial differential equations (PDEs) for an effective and robust numerical solution is proposed and used to convert the model into a system of ordinary differential equations (ODEs). 
\rev{A sensitivity analysis with respect to input current sampling time,  positive and negative particles radial discretization points,  and solver tolerances is performed.  Given the uncertainty of some model parameters,  the sensitivity analysis is performed for different realizations of such parameters,  and the best combination  of solver tolerances, discretization points,  and input current sampling time is determined in a newly introduced probabilistic framework. } 
This allows to obtain a fast solution of the governing equations while guaranteeing accuracy and numerical stability.  Moreover,  an optimization problem is formulated to identify the unknown model parameters under different C-rate.  Ultimately, the proposed approach,  while accounting for both solid phase and electrolyte dynamics,  allows to reduce the computational burden with respect to a P2D model \cite{li2015modeling}.  

The remainder of the paper is organized as follows.  Section \ref{sec:physics} describes the physical principles related to intercalation and deintercalation in $\mathrm{LiFePO_4}$/graphite batteries. In Section \ref{sec:governingeq},  the core-shell ESPM governing equations are formulated.  In Section \ref{sec:numericalsol},  core-shell model equations are discretized and converted into a system of ODEs.  \rev{Moreover,  a convenient state-space representation is provided that, when compared to the traditional ESPM state space formulation, contains a new state for the moving core-shell boundary.  In Section \ref{sec:solversens},  results of the sensitivity analysis of numerical solutions are shown (details can be found in Appendixes A and B).  Section \ref{sec:paramid} describes the parameter identification procedure.  Finally,  in Section \ref{sec:results} the model performance over charge and discharge experimental data is shown.}

\section{Physical principles}\label{sec:physics}
\begin{figure}[!t]
\centering
\includegraphics[width = 0.5\columnwidth]{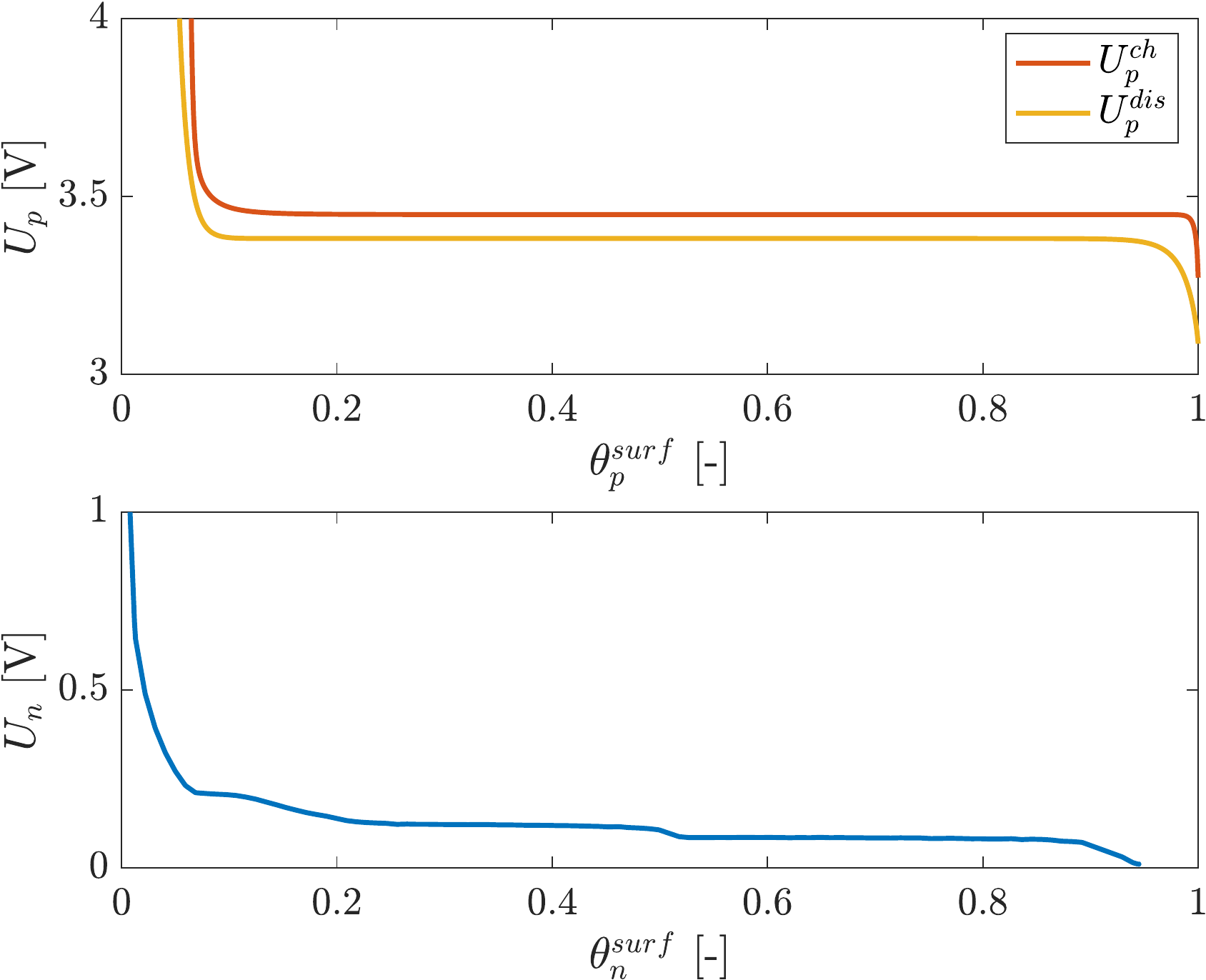}
\caption{Positive electrode charge  $U_p^{ch}$ and discharge $U_p^{dis}$ OCPs (top). Negative electrode OCP (bottom). }\label{ocps}
\end{figure}

In lithium-ion batteries,  lithium moves from the positive to the negative electrode during charging ($I<0$) and from the negative to the positive electrode during discharging ($I>0$).  During charging,  lithium leaves the positive electrode (deintercalation) and enters the negative one (intercalation).  Conversely,  during discharging, the negative electrode experiences deintercalation as lithium intercalates the positive electrode. 
For a LiFePO$_\mathrm{4}$-graphite system,  the half-cell reactions at the negative and positive electrode, are, respectively:

\rev{\begin{equation}
\textit{Negative electrode:\ \ }
\mathrm{Li_{1-a}C_6+a Li^+ +a\hspace{0.1em}e^- \xleftrightarrows[\begin{subarray}{c}\mathrm{(intercalation)}\\ \mathrm{charge}\end{subarray}]{\begin{subarray}{c}\mathrm{discharge}\\ \mathrm{(deintercalation)}\end{subarray}} LiC_6}
\end{equation}
\begin{equation}
\textit{Positive electrode:\ \ }
\mathrm{LiFePO_4 \xleftrightarrows[\begin{subarray}{c}\mathrm{(deintercalation)}\\ \mathrm{charge}\end{subarray}]{\begin{subarray}{c}\mathrm{discharge}\\ \mathrm{(intercalation)}\end{subarray}}  Li_{1-b}FePO_4 + \mathrm{b}Li^+ + \mathrm{b}\hspace{0.1em}e^-}
\end{equation}
where $\mathrm{a}$ and $\mathrm{b}$ are coefficients defined between 0 and 1.  According to  \cite{Yamada2005},  during the $\mathrm{LiFePO_4/FePO_4}$  phase transition  most of the positive electrode reaction is dominated by the coexistence of two phases:
\begin{itemize}
\item $\alpha$-phase:  Li-poor phase $\mathrm{Li_{\alpha}FePO_4}$,  with  $\alpha=1-\mathrm{b}$ and $\alpha\simeq0$;
\item $\beta$-phase: Li-rich phase $\mathrm{Li_{\beta}FePO_4}$,  with  $\beta=1-\mathrm{b}$ and $\beta\simeq1$.
\end{itemize}
$\alpha$-phase and $\beta$-phase are defined in terms of the positive particle normalized lithium concentration (i.e.,  the ratio between the actual and maximum lithium concentration):
\begin{equation}
\theta_{p}^{\alpha} = {c_{s,p}^{\alpha}}/{c_{s,p}^{max}}, \quad \theta_{p}^{\beta} = {c_{s,p}^{\beta}}/{c_{s,p}^{max}}
\label{eq:neweqnormconc}
\end{equation} 
where $c_{s,p}^{\alpha}$ and $c_{s,p}^{\beta}$ are the lithium concentrations for $\alpha$-phase and $\beta$-phase (assumed uniform in the electrode),  and $c_{s,p}^{max}$  is the maximum concentration.}
\noindent A pictorial representation of the positive electrode with the two phases (\textit{Phase \#1} and \textit{Phase \#2}) is shown in Figure \ref{fig:battespm}(a).  The meaning of \textit{Phase \#1} and \textit{Phase \#2} changes depending on whether the battery is being charged or discharged.   \rev{Furthermore,  two open circuit potentials  (OCPs) for the positive electrode are defined for the  charge and discharge operations.  In this work,  the OCPs are provided by the industrial partner of the project (see, Figure~\ref{ocps}).  The negative electrode OCP is obtained from the Galvanostatic Intermittent Titration Technique (GITT) test \cite{gitt}.  Positive electrode OCPs -- $U_p^{ch}$ and $U_p^{dis}$ in Equation \eqref{eq:cell_volt_5} -- are obtained fitting charge and discharge experimental OCPs measured at low C-rate (C/50).  Experiments for positive and negative electrode's OCPs are performed on the half-cell with lithium metal electrode as a reference.}

During discharge (see, Figure  \ref{fig:cs_1}(a)),  lithium intercalates into the positive electrode and the $\alpha$-phase is formed.  This corresponds to a rapid decrease of the OCP.  As discharge continues,  lithium concentration increases in the positive electrode and, once the normalized concentration $\theta_p^\alpha$ is reached,  the formation of the $\beta$-phase starts.  Here it is when  the $\alpha$-phase starts transitioning into the $\beta$-phase and the coexistence of two phases leads to a constant OCP (Figure  \ref{fig:cs_1}(b)).  The transition ends at the normalized concentration $\theta_p^\beta$.  After this point,  the positive particle is all in its $\beta$-phase and the OCP decreases until the end of discharge is reached (Figure \ref{fig:cs_1}(c)).   As shown in Figure \ref{fig:cs_2},  during charge,  the process is reversed and the positive particle transitions from $\beta$-phase (Figure \ref{fig:cs_2}(c)) to $\alpha$-phase (Figure \ref{fig:cs_2}(a)).  As shown in \cite{dreyer2010thermodynamic},  at the start of the charging process the surface of the positive electrode is in $\beta$-phase,  with a corresponding different OCP.  Ultimately,  the presence of two OCPs for the positive electrode is related to the voltage path dependence behavior.

\section{Cell governing equations}\label{sec:governingeq}
\subsection{Model development}\label{sec:model_dev}
\begin{figure}[!t]
\centering
\includegraphics[width = 0.65\columnwidth]{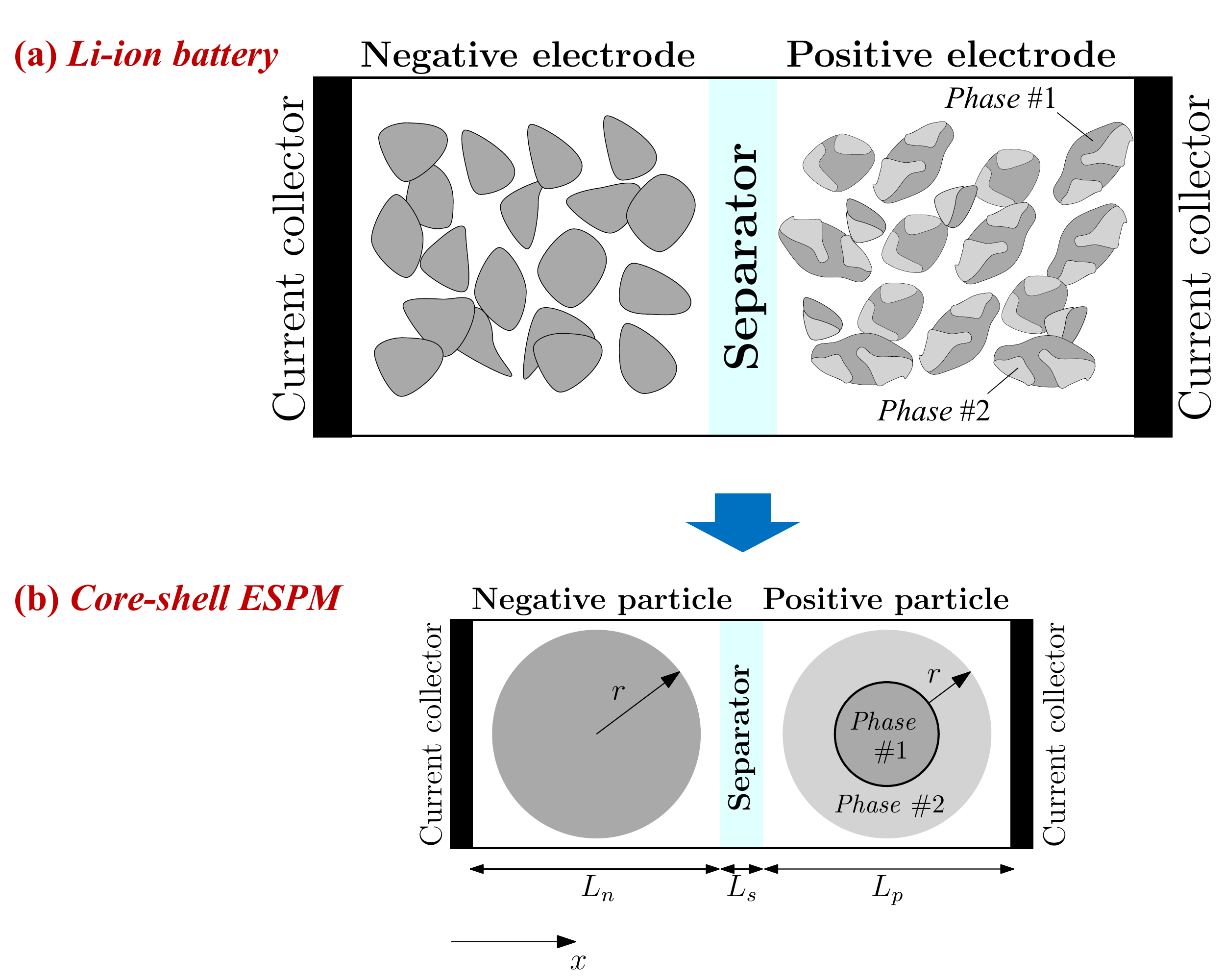}
\caption{Lithium-ion battery schematic (a).  Electrodes are composed of multiple particles with different shapes and sizes.  \textit{Phase\#1} and \textit{Phase\#2} are created during intercalation/deintercalation in the $\mathrm{LiFePO_4}$ positive electrode.  Both positive and negative electrodes are modeled with a single spherical particle (b).  The positive particle is characterized by two phases: \textit{Phase} \#1 (the core) and \textit{Phase} \#2 (the shell).  Cartesian ($x$) and radial  ($r$) coordinates are shown along with the thicknesses $L_n$,  $L_s$,  and $L_p$ for negative particle, separator, and positive particle, respectively.}
\label{fig:battespm}
\end{figure}

As shown in Figure \ref{fig:battespm}(b),  the negative electrode is approximated as a single spherical particle and following  \cite{allam2020online}, electrochemical dynamics are described by means of mass transport in the electrolyte and solid phase (Equations  \eqref{eq:eq_ne1} and \eqref{eq:eq_ne3}),  and charge transport in the electrolyte phase (Equation \eqref{eq:eq_ne2}).  Mass and charge transport equations are used to model the electrolyte dynamics in the separator (Equations \eqref{eq:eq_s1} and \eqref{eq:eq_s2}). 

As shown in Figure \ref{fig:battespm}(b),  the positive electrode is modeled as one single particle with two phases: \textit{Phase} \#1 (the core) and \textit{Phase} \#2 (the shell).  Conventional mass and charge transport equations are applied to the electrolyte phase (Equations \eqref{eq:eq_pe1} and \eqref{eq:eq_pe2}).  To describe the presence of two phases,  the solid phase mass transport equation is enhanced to account for both the one-phase and two-phase regions.  As described in Section \ref{sec:physics},  the positive electrode behavior is a function of charge and discharge conditions,  that are characterized by two different OCPs: $U_p^{ch}$ and $U_p^{dis}$. \\
\noindent During \textit{discharge},  the $U_p^{dis}$ OCP is used and the positive particle is  $\alpha$-phase.
The OCP decreases until the normalized concentration reaches $\theta_p^\alpha$ and, after this point, the formation of the $\beta$-phase starts (Figure \ref{fig:cs_1}(b) and (e)).  
Two-phase coexist inside the particle whose behaviour is captured by the core-shell paradigm, where  the core and shell of the particle are at $\alpha$-phase and $\beta$-phase, respectively.  The core is at a constant and uniform concentration $c_{s,p}^\alpha = \theta_p^\alpha\cdot c_{s,p}^{max}$ and subject to a shrinking process, as discharge takes place, which replaces the $\alpha$-phase with $\beta$-phase.  This phenomenon is modeled by means of the following mass balance equation:
\begin{equation}
\mathrm{sign}(I)(c_{s,p}^\alpha-c_{s,p}^\beta)\frac{d r_p}{d t} = D_{s,p}\frac{\partial c_{s,p}}{\partial r}\bigg|_{r = r_p}
\label{eq:massbal}
\end{equation}
Equation \eqref{eq:massbal} describes the motion of the interface (or boundary) $r_p$ between the $\alpha$-phase and $\beta$-phase while assuming $d r_p/d t$ to be function of the concentration gradient $\partial c_{s,p}/\partial r$ only.  In Equation \eqref{eq:massbal},  $c_{s,p}^\alpha$ and $c_{s,p}^\beta$ are constants defined as a function of $ \theta_p^\alpha$ and $ \theta_p^\beta$,  respectively \rev{(see Equation \eqref{eq:neweqnormconc})},  $c_{s,p}$ is the solid phase concentration,  $D_{s,p}$ is the constant solid phase diffusion coefficient,  and $r$ is the radial coordinate (i.e., the coordinate along the radius of the particle).  The  term $\mathrm{sign}(I)$ accounts for the fact that, during discharge,  $\alpha$-phase transitions to $\beta$-phase and that the opposite happens during charge.  In Figure \ref{fig:cs_1}(e), the moving boundary $r_p$ is the distance between the center of the particle and the interface between $\alpha$-phase and $\beta$-phase.  The model is completed by the following boundary condition:
\begin{equation}
c_{s,p}\big|_{r=r_p} = c_{s,p}^{\beta}
\label{eq:boundary}
\end{equation}
enforcing the concentration at the interface $r_p$ to be always equal to $c_{s,p}^\beta$.  Assuming the transition from Figure \ref{fig:cs_1}(d) to (e) to happen at the time instant $\bar{t}$,  the following initial conditions are introduced:
\begin{equation}
r_p\big|_{t=\bar{t}} = R_p-\epsilon
\label{eq:ic_1}
\end{equation}
\begin{equation}
c_{s,p}\big|_{t =\bar{t}~\wedge~r\in[0,R_p]}=c_{s,p}^\alpha
\label{eq:ic_2}
\end{equation}
Equation \eqref{eq:ic_1} is the initial condition for the moving boundary $r_p$ when, at time instant $\bar{t}$,  the system evolves from one-phase to two-phases.  \rev{At $\bar{t}$,  the boundary $r_p$ is initialized at $R_p-\epsilon$,  with $\epsilon$ a small enough coefficient ($<0.001R_p$)}.  Equation \eqref{eq:ic_2} defines the initial condition for core and shell solid phase concentrations at the transition time instant $\bar{t}$.  The core is at constant and uniform concentration $c_{s,p}^\alpha$ and,  as the core shrinks,  $c_{s,p}^\alpha$ is replaced by the $\beta$-phase concentration $c_{s,p}^\beta$.  In the shell region (i.e., for $r\geq r_p$),  the solid phase concentration starts at $c_{s,p}^\beta$ and rises according to the governing equation \eqref{eq:eq_pe3}.  Once the core is completely depleted,  the particle is fully in $\beta$-phase and lithium intercalates until the end of the discharge process,  as shown in Figure \ref{fig:cs_1}(f).  

During charge,  the opposite process,  described in Figure \ref{fig:cs_2},  occurs.  The core is in $\beta$-phase and shrinks until the whole particle is in $\alpha$-phase.  General equations for charge and discharge are provided in Equations \eqref{eq:eq_pe4},  \eqref{eq:currconv_2},  and \eqref{eq:currconv_3}.

Governing equations are summarized in Table \ref{table:ESPM_table_1}.  Additional equations for the core-shell ESPM\,--\,transport parameters, active area, porosity,  cell voltage, overpotential, and $SOC$\,--\,are summarized in Tables \ref{table:ESPM_table_2} and \ref{table:ESPM_table_2b}.

\subsubsection{Transition from one-phase to two-phase}\label{sec:smoothtransition}
\begin{figure}[!t]
\centering
\includegraphics[width = 0.50\columnwidth]{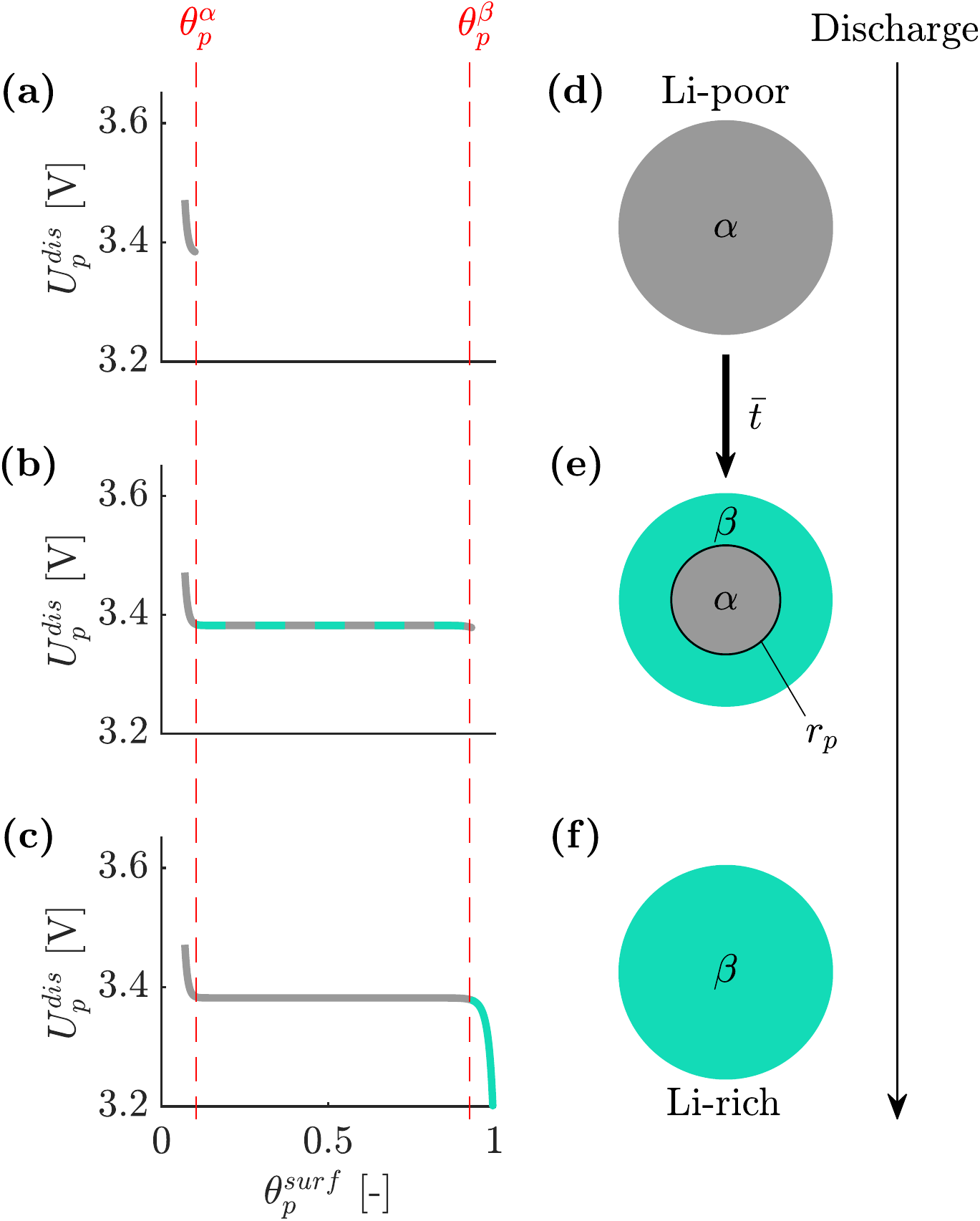}
\caption{Positive electrode OCP -- $U_p^{dis}$ -- (left) and corresponding positive particle graphical representation (right).  Figures (a), (b), and (c) show the behavior of the positive electrode OCP during discharge.  During the one-phase regions (a) and (c), the potential decreases.  As shown in (b), the coexistence of two phases leads to a flat OCP.  Figures (d), (e), and (f) show the single particle used to describe the one-phase regions ((d) and (f)) and, in particular,  the transition from the $\alpha$-phase to $\beta$-phase (e).  The transition from one-phase (d) to two-phase (e) at time $\bar{t}$ is shown.}
\label{fig:cs_1}
\end{figure}

\begin{figure}[!t]
\centering
\includegraphics[width = 0.50\columnwidth]{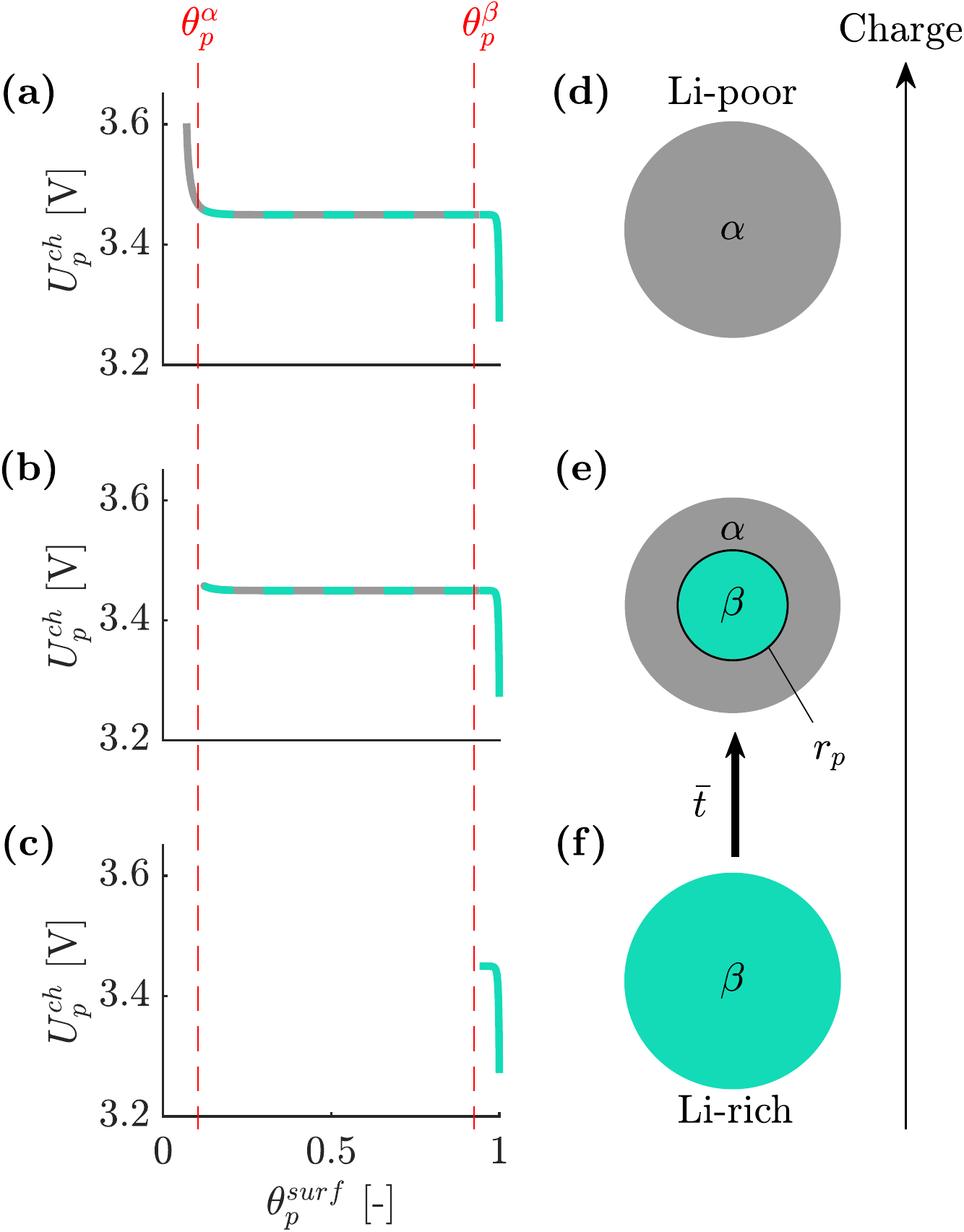}
\caption{Positive electrode OCP -- $U_p^{ch}$ -- (left) and corresponding particle graphical representation (right).  Figures (c), (b), and (a) show the behavior of the positive electrode OCP during discharge.  During the one-phase regions (c) and (a), the potential increases.  As shown in (b), the coexistence of two phases leads to a flat OCP.  Figures (f), (e), and (d) show the single particle used to describe the one-phase regions ((f) and (d)) and, in particular,  the transition from the $\beta$-phase to $\alpha$-phase (e).  The transition from one-phase (f) to two-phase (e) at time $\bar{t}$ is shown.}
\label{fig:cs_2}
\end{figure}

One-phase concentration dynamics are modeled according to Equations \eqref{eq:eq_pe3} and \eqref{eq:eq_pe5}.  After the transition to the two-phase region,  the core is assumed to be at constant and uniform concentration -- i.e., $c_{s,p}^\alpha$ and $c_{s,p}^\beta$ (see,  Figures \ref{fig:cs_1}(e) and \ref{fig:cs_2}(e), respectively) -- while the shell concentration varies according to Equations \eqref{eq:eq_pe3} and \eqref{eq:eq_pe4}.    

The transition between one-phase (Figures \ref{fig:cs_1}(d) and \ref{fig:cs_2}(f)) and two-phase (Figures \ref{fig:cs_1}(e) and \ref{fig:cs_2}(e)) is crucial and must ensure mass conservation.  
The positive particle bulk normalized lithium concentration $\theta_{p}^{bulk}$ is computed as follows:
\begin{equation}
\theta_{p}^{bulk}  = \frac{3}{c_{s,p}^{max}R_p^3}\int_{0}^{R_p}c_{s,p}r^2dr
\label{eq:nconc_1}
\end{equation}
The transition from the one-phase to two-phase is performed when $\theta_{p}^{bulk} = \theta_{p}^{*}$, i.e., when the following mass balance is satisfied:
\begin{equation} 
\frac{3}{c_{s,p}^{max}R_p^3}\int_{0}^{R_p}c_{s,p}r^2dr = \frac{c_{c,p}^*}{c_{s,p}^{max}}\ \longrightarrow\ 
\int_{0}^{R_p}c_{s,p}r^2dr = c_{s,p}^{*}\frac{R_p^3}{3}
\label{eq:nconc_2}
\end{equation}
with $c_{s,p}$ the one-phase concentration at the transition time instant $\bar{t}$.  In Equation \eqref{eq:nconc_2},  the normalized core concentration $\theta_{p}^{*}$ is equal to $\theta_{p}^\alpha$ or $\theta_{p}^\beta$ depending on whether the battery is being discharged (Figure \ref{fig:cs_1}) or charged (Figure \ref{fig:cs_2}).  Similarly, in Equation \eqref{eq:nconc_2},  $c_{s,p}^*$ is equal to $c_{s,p}^\alpha$ and $c_{s,p}^\beta$ for discharge and charge, respectively.

The transition from two-phase (Figures \ref{fig:cs_1}(e) and \ref{fig:cs_2}(e)) to one-phase (Figures \ref{fig:cs_1}(f) and \ref{fig:cs_2}(d)) is already taken into account by the core-shell formulation, i.e., the transition happens when the core is completely depleted and $r_p$ reaches zero.  

\section{Numerical solution}\label{sec:numericalsol}
The system of coupled PDEs in Table \ref{table:ESPM_table_1} is discretized using the finite difference method (FDM),  for mass transport in the solid phase,  and finite volume method (FVM), for mass transport in the electrolyte phase.  This allows to convert the governing equations of the core-shell ESPM model into a system of ODEs and algebraic equations. The system of ODEs can be rewritten into a convenient state-space representation and solved relying on the \textsc{Matlab} numerical solver \texttt{ode15s} \cite{mathworks2000matlab}.  This solver is effective for nonlinear/stiff problems and suitable for the solution of electrochemical transport equations.  

In this work, we focus on the numerical solution of the positive electrode solid phase mass transport dynamics,  where the core-shell process takes place.  In the next sections,  Equations \eqref{eq:eq_pe3}  and \eqref{eq:eq_pe4}  in Table \ref{table:ESPM_table_1} are discretized relying on the FDM \cite{takahashisub}.  The discretization of negative electrode solid phase and electrolyte mass transport dynamics is provided in Table \ref{tab:discradditional}.  For additional details,  readers are referred to \cite{weaver2020novel}.  

\subsection{Coordinate system transformation: normalization}
Starting from the positive electrode core-shell model in Section \ref{sec:governingeq}, the transformation proposed in \cite{srinivasan2004discharge} is used to move from the radial coordinate system ($r$) to the normalized coordinate $\chi$:
\begin{equation}
\chi = \frac{r - r_p}{R_p - r_p} \in [0,1]
\label{eq:changecoord}
\end{equation}
where $r$ represents a radial position in the particle and $r_p$ is the moving boundary. This transformation remaps the discretization of the shell region from $[r_p,R_p]$ to $[0,1]$,  making the domain stationary while the boundary is moving.  \rev{In Equation  \eqref{eq:ic_1},  the small enough $\epsilon$ avoids the rise of singularities during transitions from one-phase to two-phase.}

As proposed in \cite{li2015modeling}, the left hand side of the positive particle diffusion Equation \eqref{eq:eq_pe3} is transformed from $r$  to $\chi$ domain as follows:
\begin{equation}
\left(\frac{\partial c_{s,p}}{\partial t}\right)_r = \frac{\partial c_{s,p}}{\partial \chi}\frac{\partial \chi}{\partial t} + \left(\frac{\partial c_{s,p}}{\partial t}\right)_\chi
\label{eq:term_1}
\end{equation}
Introducing the following relationships:
\begin{equation}
\frac{\partial\chi}{\partial t} = \frac{\partial \chi}{\partial r_p}\frac{\partial r_p}{\partial t},\quad \frac{\partial \chi}{\partial r_p} = \frac{\chi-1}{R_p-r_p}
\end{equation}
Equation \eqref{eq:term_1} can be rewritten as:
\begin{equation}
\left(\frac{\partial c_{s,p}}{\partial t}\right)_r = \frac{\partial c_{s,p}}{\partial \chi} \left(\frac{\chi-1}{R_p-r_p}\right)\frac{\partial r_p}{\partial t} + \left(\frac{\partial c_{s,p}}{\partial t}\right)_\chi
\label{eq:part_1}
\end{equation}
The change of coordinates applied to the right hand side of Equation \eqref{eq:eq_pe3} is obtained using the following transformations:
\begin{equation}
\begin{split}
&\frac{\partial^2 c_{s,p}}{\partial r^2} = \frac{\partial}{\partial r}\left(\frac{\partial c_{s,p}}{\partial r}\right) = \frac{\partial}{\partial r}\left(\frac{\partial c_{s,p}}{\partial \chi}\frac{\partial\chi}{\partial r}\right)\\
 \frac{\partial c_{s,p}}{\partial r}&=\frac{\partial c_{s,p}}{\partial \chi}\frac{\partial\chi}{\partial r}, \quad\frac{\partial}{\partial r} = \frac{\partial\chi}{\partial r}\frac{\partial}{\partial\chi}, \quad \frac{\partial\chi}{\partial r} = \frac{1}{R_p-r_p}
\end{split}
\label{eq:changecoord_2}
\end{equation}
Given Equation \eqref{eq:changecoord_2},  the following reformulation is obtained:
\begin{equation}
D_{s,p}\frac{\partial^2 c_{s,p}}{\partial r^2} = \frac{\partial^2 c_{s,p}}{\partial \chi^2}\frac{D_{s,p}}{(R_p-r_p)^2}
\label{eq:part_2}
\end{equation}
\begin{equation}
\frac{2 D_{s,p}}{r}\frac{\partial c_{s,p}}{\partial r} = \frac{\partial c_{s,p}}{\partial\chi}\left[\frac{2D_{s,p}}{r(R_p-r_p)}\right]
\label{eq:part_3}
\end{equation}
Given Equations \eqref{eq:part_1},  \eqref{eq:part_2},  and \eqref{eq:part_3},  the solid phase mass transport in the positive particle (Equation \eqref{eq:eq_pe3}) is rewritten as:
\begin{equation}
\frac{\partial c_{s,p}}{\partial t} = \frac{\partial^2 c_{s,p}}{\partial\chi^2}\left[\frac{D_{s,p}}{(R_p-r_p)^2}\right]+\frac{\partial c_{s,p}}{\partial\chi}\left[\frac{2D_{s,p}}{r(R_p-r_p)}\right]-\frac{\partial c_{s,p}}{\partial\chi}\frac{\partial r_p}{\partial t}\left[\frac{\chi-1}{R_p-r_p}\right]
\label{eq:final_1}
\end{equation}

Starting from Equation \eqref{eq:changecoord_2},  boundary conditions and mass balance in Equation \eqref{eq:eq_pe4} are  rewritten in terms of the $\chi$ coordinate:  
\begin{equation}
\frac{\partial c_{s,p}}{\partial \chi}\bigg\vert_{\chi=1} = \frac{I(R_p-r_p)}{D_{s,p}a_{p}A_{cell}FL_p}
\label{eq:final_2}
\end{equation}
\begin{equation}
c_{s,p}\big|_{\chi=0} = \mathrm{g}(I) \quad c_{s,p}\big|_{t=\bar{t}} = \mathrm{ic}_k
\label{eq:final_3}
\end{equation}
\begin{equation}
\mathrm{sign}(I)(c_{s,p}^\alpha - c_{s,p}^\beta)(R_p-r_p)\frac{d r_p}{d t} = D_{s,p}\frac{\partial c_{s,p}}{\partial \chi}\bigg|_{\chi=0} 
\label{eq:final_4}
\end{equation}
where $\chi = 0$ coincides with the moving boundary $r_p$,  $\chi=1$ corresponds to the surface or $R_p$, and $g(I)$ is the concentration at the moving boundary (as defined in Equation \eqref{eq:currconv_2} of Table \ref{table:ESPM_table_2}).

\subsection{Discretization}
\begin{figure}[!t]
\centering
\includegraphics[width = 0.5\textwidth]{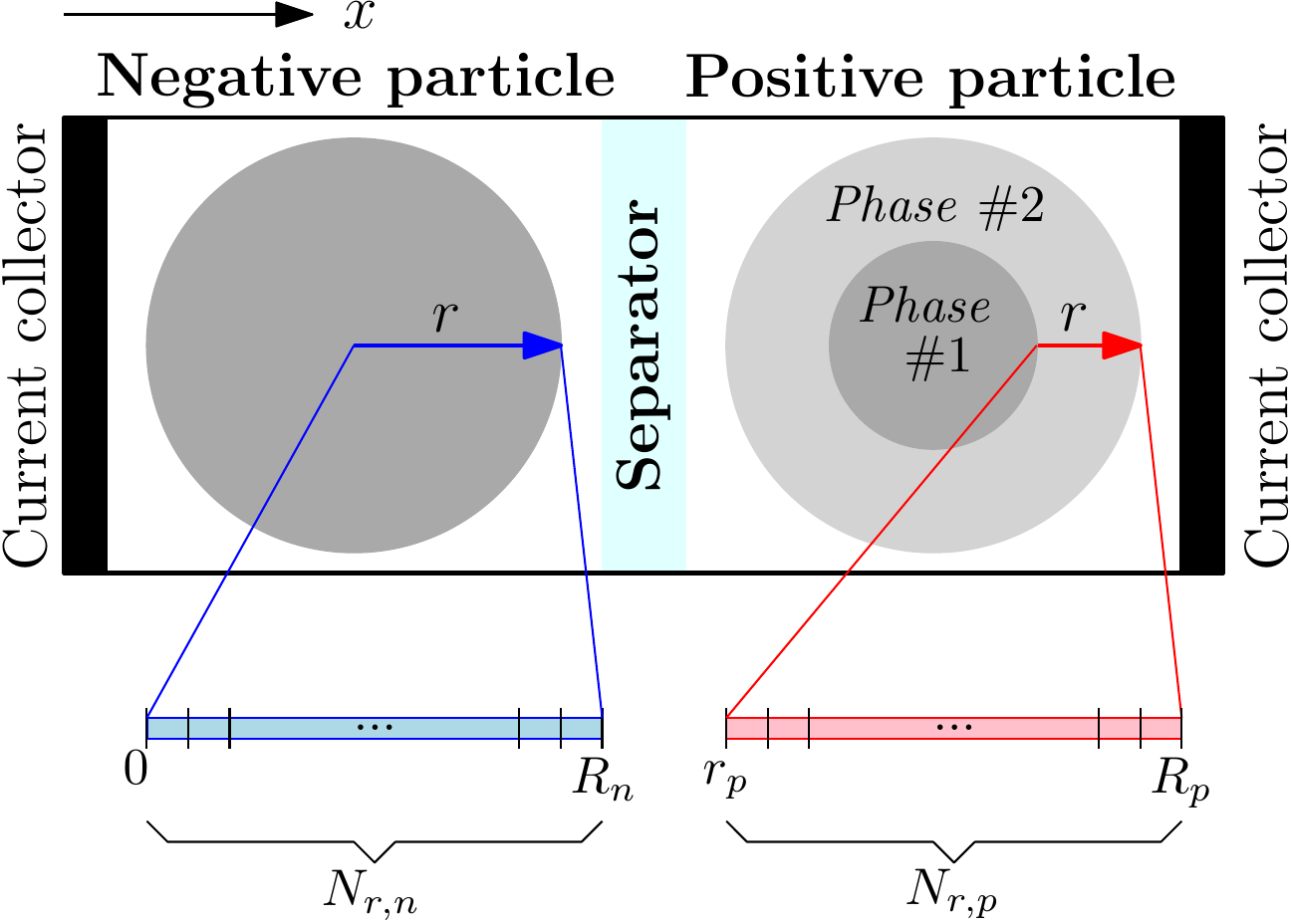}
\caption{Core-shell ESPM: discretization of the negative and positive particles.  For the positive particle, the discretization grid is varying as a function of the moving boundary $r_p$.}
\label{fig:discr_cs}
\end{figure}

The core-shell framework -- Equations \eqref{eq:final_1},  \eqref{eq:final_2}, \eqref{eq:final_3},  and \eqref{eq:final_4} -- is discretized into $N_{r,p}$ points \rev{(Figure \ref{fig:discr_cs})}.  The right-sided and central finite difference schemes are used for the first and second derivative approximations:
\begin{equation}
\frac{\partial u}{\partial \chi} \bigg\vert_{\chi_l} \approx \frac{u_{l+1} -  u_l}{\Delta_\chi} 
\end{equation}
\begin{equation}
\frac{\partial^2 u}{\partial \chi^2} \bigg\vert_{\chi_l} \approx \frac{u_{l+1} -  2u_l + u_{l-1}}{\Delta_\chi^2}
\end{equation}
where $u$ is a generic variable and $l$ defines the index of the discretization point $\chi_l$:
\begin{equation}
\chi_l = \frac{r_l-r_p}{R_p-r_p},\ \Delta_\chi = \chi_l - \chi_{l-1}
\label{eq:chiremap}
\end{equation}
In Equation \eqref{eq:chiremap},  $r_l$,  the radial position along the shell region,  takes the following form:
\begin{equation}
r_l=r_p+l\Delta_r, 	\ \Delta_r = \frac{R_p-r_p}{N_{r,p}-1}
\end{equation}

\subsubsection{Discretization of mass balance}
Equation \eqref{eq:final_4} is rewritten as:
\begin{equation}
\frac{d r_p}{d t} = \frac{\mathrm{sign}(I)D_{s,p}}{(c_{s,p}^\alpha- c _{s,p}^\beta)(R_p-r_p)}\frac{\partial c_{s,p}}{\partial\chi}\bigg\vert_{0}
\end{equation}
Approximating the term $\frac{\partial c_{s,p}}{\partial\chi}\big\vert_{0}$ as:
\begin{equation}
\frac{c_{s,p}}{\partial\chi}\bigg\vert_{0}\approx \frac{c_{s,p_1}-c_{s,p_0}}{\Delta_\chi}
\end{equation}
and given that  $c_{s,p_{0}} = \mathrm{g}(I)$ (from Equation \eqref{eq:final_3}),  the discretized version of the mass balance is obtained:
\begin{equation}
\begin{split}
&\frac{d r_p}{d t} = \frac{M_1}{\Delta_\chi}(c_{s,p_1}-\mathrm{g}(I)) \\
& M_1 = \frac{\mathrm{sign}(I)D_{s,p}}{(c_{s,p}^\alpha - c_{s,p}^\beta)(R_p-r_p)}
\end{split}
\label{eq:discr_3}
\end{equation}

\subsubsection{Discretization of boundary conditions}\label{sec:nextpar}
At $\chi = 0$,  the moving boundary condition in Equation \eqref{eq:final_3} takes the following form:
\begin{equation}
c_{s,p_{0}} = \mathrm{g}(I) \rightarrow \frac{\partial c_{s,p_0}}{\partial t} = 0
\label{eq:discr_1}
\end{equation}
and the fixed boundary condition \eqref{eq:final_2} is rewritten as:
\begin{equation}
\frac{\partial c_{s,p}}{\partial \chi}\bigg\vert_{N_{r,p}-1} = \frac{I(R_p-r_p)}{D_{s,p}a_{p}A_{cell}FL_p}
\label{eq:fb_disc}
\end{equation}
Approximating the left hand side of Equation \eqref{eq:fb_disc} as:
\begin{equation}
\frac{\partial c_{s,p}}{\partial \chi}\bigg\vert_{N_{r,p}-1} \approx \frac{c_{s,p_{N_{r,p}}}-c_{s,p_{N_{r,p}-1}}}{\Delta_\chi},
\end{equation}
the following is obtained:
\begin{equation}
\begin{split}
& c_{s,p_{N_{r,p}}} = c_{s,p_{N_{r,p}-1}} + M_2I \\
& M_2 = \frac{(R_p-r_p)\Delta_\chi}{D_{s,p}a_pA_{cell}FL_p}
\end{split}
\label{eq:discr_2}
\end{equation}

\subsubsection{Discretization of the solid phase mass transport}
Equation \eqref{eq:final_1} is discretized relying on both the right-sided and central finite difference schemes. For $l \in [1,N_{r,p}-2]$, the following expression is obtained:
\begin{equation}
\frac{\partial c_{s,p}}{\partial t}\bigg\vert_{l} = \frac{M_3}{\Delta_\chi^2}(c_{s,p_{l+1}}-2c_{s,p_l} +c_{s,p_{l-1}}) + \frac{M_4}{\Delta_\chi}(c_{s,p_{l+1}}-c_{s,p_l})
\label{eq:discr_4}
\end{equation}
with $M_3$ and $M_4$ defined as:
\begin{equation}
\begin{split}
& M_3 = \frac{D_{s,p}}{(R_p-r_p)^2}\\
& M_4 = \frac{2D_{s,p}}{[\chi_l(R_p-r_p)+r_p](R_p-r_p)}-\frac{\chi_l-1}{R_p-r_p}\frac{M_1}{\Delta_\chi}(c_{s,p_1}-\mathrm{g}(I))
\end{split}
\end{equation}
At $l=N_{r,p}-1$,  the discretized solid phase mass transport equation takes the following form:
\begin{equation}
\frac{\partial c_{s,p}}{\partial t}\bigg\vert_{N_{r,p}-1} = \frac{M_3}{\Delta_\chi^2}(c_{s,p_{N_{r,p}}}-2c_{s,p_{N_{r,p}-1}} +c_{s,p_{N_{r,p}-2}}) + \frac{M_4}{\Delta_\chi}(c_{s,p_{N_{r,p}}}-c_{s,p_{N_{r,p}-1}})
\label{eq:tmp_discr_fick}
\end{equation}
From Equation \eqref{eq:discr_2},  Equation \eqref{eq:tmp_discr_fick} is rewritten as:
\begin{equation}
\frac{\partial c_{s,p}}{\partial t}\bigg\vert_{N_{r,p}-1} = \frac{M_3}{\Delta_\chi^2}(M_2I-c_{s,p_{N_{r,p}-1}} +c_{s,p_{N_{r,p}-2}}) + \frac{M_2M_4}{\Delta_\chi}I
\label{eq:discr_5}
\end{equation}
It is worth mentioning that, according to Equation \eqref{eq:discr_1}, at $l=0$ the time derivative $\frac{\partial c_{s,p_0}}{\partial t} = 0$.

\subsection{State-space representation}\label{sec:ssposel}
\rev{Equations \eqref{eq:discr_3},  \eqref{eq:discr_1}, \eqref{eq:discr_4}, and \eqref{eq:discr_5} constitute a system of coupled ODEs  which can be rewritten in state-space form.  This is a practical reformulation for the implementation of the core-shell governing equations and their numerical solution.  

Given the vector of discretized solid phase concentration states:
\begin{equation}
\begin{split}
&\mathbf{c}_{s,p} = [c_{s,p_1}\ c_{s,p_2}\ \dots\ c_{s,p_{N_{r,p}-2}}\ c_{s,p_{N_{r,p}-1}}]^T \in \mathbb{R}^{(N_{r,p}-1)\times 1}\\
&c_{s,p}^{surf} = c_{s,p_{N_{r,p}-1}} \in \mathbf{c}_{s,p}
\end{split}
\end{equation}
and the moving boundary $r_p$,  the state vector $\mathbf{x}$ is defined as:
\begin{equation}
\mathbf{x} = \begin{bmatrix} r_p\\ \mathbf{c}_{s,p}\end{bmatrix}\in \mathbb{R}^{N_{r,p}\times 1}\\
\label{eq:ssrepvec}
\end{equation}
Introducing the variables:
\begin{equation}
\eta_1 = \frac{M_3}{\Delta_\chi^2},\quad \eta_2 = \frac{M_4}{\Delta_\chi}, \quad \eta_3 = \frac{M_2}{\Delta_\chi}\left(M_4+\frac{M_3}{\Delta_\chi}\right),\quad \eta_4 = \frac{M_1}{\Delta_\chi}
\end{equation}
the positive particle state-space representation is given by:
\begin{equation}
\dot{\mathbf{x}} = \eta_1\mathbf{A}_{s,p1}\mathbf{x} + \eta_2\mathbf{A}_{s,p2}\mathbf{x} + \eta_3\mathbf{B}_{s,p}I+\eta_1\mathbf{G}_{s,p}
\label{eq:sscore}
\end{equation}
In Equation \eqref{eq:sscore}, matrices $\mathbf{A}_1$ and $\mathbf{A}_2$ take the following expressions:
\begin{equation}
\mathbf{A}_{s,p1} = 
\begin{bmatrix}
0 & \eta_4/\eta_1  & 0 & 0 & 0 & \dots & 0  \\
0 & -2 & 1 & 0 & 0 & \dots & 0  \\
0 &1 & -2 & 1 & 0 & \dots & 0 \\
0 &0 & 1 & -2 & 1 & \dots & 0 \\
0 &0 & 0 & 1 & -2 & \dots & 0 \\
\vdots & \vdots & \vdots & \vdots & \vdots & \ddots & \vdots\\
0& 0 & 0 & 0 & 0 & \dots &  -1\\
\end{bmatrix}_{N_{r,p}\times N_{r,p}}
\end{equation}
\begin{equation}
\mathbf{A}_{s,p2} = 
\begin{bmatrix}
0 & 0 & 0 & 0 & 0 & \dots & 0 \\
0 & -1 & 1 & 0 & 0 & \dots & 0 \\
0 & 0 & -1 & 1 & 0 & \dots & 0 \\
0 &0 & 0 & -1 & 1 & \dots & 0 \\
0 &0 & 0 &  0 & -1 & \dots & 0 \\
\vdots & \vdots & \vdots & \vdots & \vdots & \ddots & \vdots \\
0 &0 & 0 & 0 & 0 & \dots & 0 \\
\end{bmatrix}_{N_{r,p}\times N_{r,p}}
\end{equation}
Similarly, vectors $\mathbf{B}$ and $\mathbf{G}$ are defined as:
\begin{equation}
\mathbf{B}_{s,p} = 
\begin{bmatrix}
0\\
0\\
0\\
\vdots\\
0\\
1\\
\end{bmatrix}_{N_{r,p}\times 1}, \quad \mathbf{G}_{s,p} = 
\begin{bmatrix}
-\eta_4/\eta_1\mathrm{g}(I)\\
\mathrm{g}(I)\\
0\\ 
\vdots\\
0\\
0\\
\end{bmatrix}_{N_{r,p}\times 1}
\end{equation}
Compared to a conventional ESPM state-space representation, \cite{weaver2020novel},  the state vector $\mathbf{x}$ in Equation \eqref{eq:ssrepvec} has the moving boundary $r_p$ as the new state from which the positive particle solid phase concentration depends on.  Overall,  the positive particle is described by $N_{r,p}$ states,  among which $N_{r,p}-1$ are associated with the positive particle solid phase concentration and $1$ with the moving boundary.  This is in contrast with the conventional ESPM formulation,  where the discretization leads to $N_{r,p}$ states describing the positive particle solid phase concentration dynamics. 

State-space representations for negative electrode solid phase and electrolyte mass transport dynamics,  derived as shown in \cite{weaver2020novel},  are summarized in Table \ref{tab:discradditional}.  The electrolyte concentration ($c$) is discretized along the Cartesian coordinate $x$ with $N_{x,n}$,  $N_{x,s}$,  and $N_{x,p}$ discretization points for negative particle,  separator,  and positive particle,  respectively.  The negative particle  solid phase concentration ($c_{s,n}$) is discretized along the radial coordinate $r$,  with $N_{r,n}$ discretization points defined between the radial coordinates 0 and $R_n$ (Figure \ref{fig:discr_cs}). }

\section{Sensitivity analysis of numerical solutions}\label{sec:solversens}
\rev{Previous works from the authors \cite{allam2020online,weaver2020novel,pozzato2021modeling} show that a number of discretization points $N_{x,n} = N_{x,s} = N_{x,n} =10$ and $N_{r,n} = N_{r,p} = 10$ generates a stable and accurate solution of the ESPM governing equations.  In the core-shell ESPM, though, the presence of the moving boundary $r_p$ as an additional state is a potential source of numerical instabilities and lack of convergence of the solver.  A sensitivity analysis is therefore conducted in this work to find suitable solver settings for a robust solution of the ODEs system.}

\rev{The sensitivity analysis is carried out with respect to the following parameters:  the  radial coordinate discretization points ($N_{r,n}$ and $N_{r,p}$),  the input current sampling time ($dt$),  and the solver absolute and relative tolerances ($\mathrm{abstol}$ and $\mathrm{reltol}$).  Sampling time and solver tolerances control the solver convergence, whereas  the radial coordinate discretization points control the accuracy of the solution of the solid phase diffusion (i.e., the number of ODEs).  In this work,  $N_{r,n} = N_{r,p} = N_r$ and results of the sensitivity analysis are shown as a function of $N_r$.  On the other hand, for the  solution of the electrolyte phase concentration dynamics along the Cartesian coordinate, we consider  $N_{x,p}=N_{x,s}=N_{x,n}=10$,  as these parameters do not affect the solver stability.

Different combinations of solver settings $ \mathcal{C}=\{N_r,dt,\mathrm{reltol},\mathrm{abstol}\}$ were tested and the optimal combination of values, chosen based on a probabilistic framework explained in details  in Appendixes A and B,  is given by:
\begin{equation}
\hat{\mathcal{C}} = \{70,50,1\times 10^{-5},\mathrm{reltol}\times 0.001\}
\label{eq:solverbest}
\end{equation} 
The optimal solver settings in Equation \eqref{eq:solverbest} are used for all identifications and simulations shown in the next sections.}

\section{Parameters identification}\label{sec:paramid} 
In this work,  the values of region thicknesses ($L_p$, $L_n$,  $L_s$),  maximum solid phase concentrations ($c_{s,p}^{max}$,  $c_{s,n}^{max}$),  porosity of electrodes and separator ($\varepsilon_p$, $\varepsilon_n$, $\varepsilon_s$),  active volume fractions ($\nu_p$, $\nu_n$, $\nu_s$),  and transference number ($t_+$) were provided by the industrial partner of the project.  \gbrev{Readers can refer to \cite{prada2013simplified} to find similar values for maximum solid phase concentrations,  transference number,  porosities of separator and negative electrode,  and active volume fractions of separator and negative electrode.  Porosity and active volume fraction of the positive electrode,  and region thicknesses can be separately identified,  for example,  following the methodology in \cite{arun2019}.} 

Unknown model parameters are identified using voltage \textit{vs} capacity data for a $\mathrm{LiFePO}_4$/graphite pouch cell charged and discharged at C/12, C/10, C/6, and C/3 constant current at 25$^\circ$C.  \rev{Charge and discharge capacities at the different C-rate are computed from experimental data and listed in Table \ref{tab:dischcapacities}.  A second cell, with the same technical specifications and tested under the same conditions,  is used to verify the goodness of the identified parameters.  Using the particle swarm optimization (PSO) algorithm\footnote{\textsc{Matlab} \texttt{particleswarm} function: \url{https://www.mathworks.com/help/gads/particleswarm.html}},  the parameter vector $\Theta$,  described in the next section,  is identified. }

The identification is performed following the line of \cite{allam2020online} and \cite{pozzato2021modeling}, i.e.,  minimizing the following cost function in constant current charge and discharge conditions:
\begin{equation}
\begin{split}
J_k\big(\Theta,{\mathcal{C}}\big) &=  \overbrace{w_1 \sqrt{\frac{1}{N} \sum_{j=1}^N\left(\frac{V_{exp}^k(j) - V^k(\Theta,{\mathcal{C}};j)}{V_{exp}^k(j)}\right)^2}}^{J_V}  + \overbrace{w_2 \sqrt{\frac{1}{N} \sum_{j=1}^N(SOC_{exp}^k(j) - SOC_n^k(\Theta,{\mathcal{C}};j))^2}}^{J_{SOC_n}} \\ &+ \underbrace{w_3 \sqrt{\frac{1}{N} \sum_{j=1}^N(SOC_{exp}^k(j) - SOC_p^k(\Theta,{\mathcal{C}};j))^2}}_{J_{SOC_p}}
\end{split}
\label{eq:costterms}
\end{equation}
where  $k\in\mathcal{K} = \{\text{charge},\text{discharge}\}$,  $\mathcal{C}$ is the solver setting (determined according to Section \ref{sec:solversens}),  $j$ is the time index,  $N$ is the number of samples,  $SOC_p^k$ and $SOC_n^k$ are the simulated state of charge at the positive and negative electrodes,  $V^k$ is the simulated voltage profile, $V_{exp}^k$ and $SOC_{exp}^k$ are the experimental cell voltage and state of charge from Coulomb counting, respectively.  The weights $w_1$,  $w_2$, and $w_3$ are user-defined dimensionless parameters here equal to one.  The parameter vector $\Theta$ is  the same for charge and discharge.

The identification problem is formulated as the following optimization problem:
\begin{flalign}
&\underset{\Theta}{\mathbf{minimize}}\hspace{10em}\ J\big(\Theta,{\mathcal{C}}\big) = \sum_{k\in \mathcal{K}} J_k\big(\Theta,{\mathcal{C}}\big)&
\label{eq:ss_ocp_1}
\end{flalign}
$\mathbf{subject\ to}$
\begin{equation*}
\begin{split}
&\mathrm{(a)\ } \text{Governing equations}\ \text{(Table \ref{tab:discradditional})}\\
&\mathrm{(b)\ }\theta^\beta_p \leq \theta_{p,0\%}\\
&\mathrm{(c)\ }\theta^\alpha_p \geq \theta_{p,100\%}\\
&\mathrm{(d)\ }\begin{cases}
	r_p^k(\Theta,{\mathcal{C}};j) \geq 0,\quad j\in[1,N-1]\\
	r_p^k(\Theta,{\mathcal{C}};N) = 0,\quad j = N
	\end{cases}
\end{split}\vspace{1em}
\end{equation*}

The minimization of the cost function is subject to the system dynamics $\mathrm{(a)}$,  and inequalities $\mathrm{(b)}$ and $\mathrm{(c)}$ to ensure the two-phase region (defined between $\theta_p^\alpha$ and $\theta_p^\beta$) to be contained inside the positive electrode stoichiometric window $\theta_{p,0\%}$-$\theta_{p,100\%}$.  Constraints $\mathrm{(d)}$ enforce the moving boundary $r_p$ to be greater than or equal to zero for $j\in[1,N-1]$.  In this work,  constant current charge and discharge profiles are considered and the cell always reaches the one-phase at the end of the charge or discharge process.  Therefore,  the moving boundary $r_p$ reaches zero for $j=N$. 

\rev{The four stoichiometric coefficients defining the stoichiometric windows in the positive and negative electrode,  respectively,  are identified once at C/12 while imposing charge conservation.  Charge and discharge capacities are computed according to the following formula:
\begin{equation}
Q_i^{k}(\Theta) = \frac{\nu_iFL_iA_{cell}c_{s,i}^{max}\left|\theta_{i,100\%}-\theta_{i,0\%}\right|}{3600}
\end{equation}
where $i\in\hat{\mathcal{M}}=\{p,n\}$ indicates the positive and negative electrode.  To enforce charge conservation,  $Q_i^{k}$ must satisfy the following constraint:\begin{equation}
\hspace{-4.5em}\mathrm{(e)\ }\underline{Q} \leq Q_i^k(\Theta) \leq \overline{Q},\quad i\in\hat{\mathcal{M}}
\end{equation}
where parameters $\overline{Q}$ and $\underline{Q}$ are suitable bounds defined as $(1+0.01)Q_{C/12}^k$ and $(1-0.01)Q_{C/12}^k$,  with $Q_{C/12}^k$ the C/12 charge and discharge capacity computed from experimental data (Table \ref{tab:dischcapacities}). These bounds are used to ensure the numerical feasibility of the identification problem.}

\subsection{Identification at C-rate C/12,  C/10,  C/6,  and C/3}\label{sec:idnum2}
\rev{Charge and discharge experimental data at C/12 are used to identify the parameter vector:
\begin{equation}
\begin{split}
\Theta_{C/12} = \big[& R_n, R_p,A_{cell},D_{s,n},D_{s,p}, \theta_{n,100\%},\theta_{n,0\%},\theta_{p,100\%},\theta_{p,0\%},\theta_p^{\alpha},\theta_p^{\beta},R_l\big]^T
\end{split}
\label{eq:idvec}
\end{equation}
where  $R_n$,  $R_p$,  and $A_{cell}$ are geometrical parameters,  $D_{s,n}$ and $D_{s,p}$ are solid phase diffusion coefficients controlling the mass transport in the negative and positive electrodes,  $\theta_{n,100\%}$, $\theta_{n,0\%}$, $\theta_{p,100\%}$, and $\theta_{p,0\%}$ define the stoichiometric window of the cell,  and $R_l$ is the lumped contact resistance.  For the implementation of the core-shell paradigm, the identification of $\theta_p^\alpha$ and $\theta_p^\beta$ is key to characterize the transition from one-phase ($\alpha$ or $\beta$) to two-phases ($\alpha$ and $\beta$).}

\rev{At C/10,  the following parameter vector is identified:
\begin{equation}
\begin{split}
\Theta_{C/10} = \big[& D_{s,n},D_{s,p}, R_l\big]^T
\end{split}
\end{equation}
Stoichiometric and geometric parameters are set as constants from the $\Theta_{C/12}$ identification results.
This is reasonable since the proposed model does not consider electrode swelling phenomena \cite{padhi1997phospho} and the geometry is not changing while increasing the C-rate.  The stoichiometric window -- defined by $\theta_{i,0\%}$ and $\theta_{i,100\%}$ -- describes the maximum and minimum concentration of lithium-ion in the positive and negative electrodes,  and it does not change while increasing the C-rate because of mass conservation.  On the other hand, the discharged (or charged) capacity at C/10 is lower than the discharged capacity at C/12.   In this situation,  the stoichiometric window is not changing, but the cell is using a narrow range inside the $\theta_{i,0\%}$ and $\theta_{i,100\%}$ window.  Therefore,  constraint (e) in the identification problem \eqref{eq:ss_ocp_1} is valid only for the C/12 scenario, where the full stoichiometric window is used,  and deactivated for the other C-rates used in the identification.  The overpotential contribution is small at C/12 and C/10,  making the parameter $k_n$ and $k_p$ not identifiable,  for the values of $k_n = k_p = 10^{-11}~[\mathrm{m^{2.5}/(mol^{0.5}s)}]$ are used.  \gbrev{These values are initial guesses that,  as done for C/6 and C/3,  should be refined through identification.}}

\rev{At C/6 and C/3,  the following parameter vector is identified:
\begin{equation}
\begin{split}
\Theta_{C/6}=\Theta_{C/3} = \big[& D_{s,n},D_{s,p}, R_l,k_n, k_p\big]^T
\end{split}
\end{equation}
In addition to the diffusion coefficients and lumped resistance -- also identified at C/10 -- since  the overpotential contribution increases at higher C-rate, the reaction rate constants $k_n$ and $k_p$ are identified. }

\rev{PSO is a nonlinear optimization algorithm that searches the optimal parameter vector inside a space defined by the parameters' upper and lower bounds and it requires  an initial guess for the parameter vector to be identified.  In this paper,  the identification is performed sequentially,  starting with C/12 and C/10, C/6 and lastly C/3  to follow.  At C/12, reasonable bounds and initial guesses are obtained from the available literature on LFP batteries \cite{li2015modeling}, \cite{li2014current}, \cite{prada2013simplified} and information provided by our industrial sponsor (Table \ref{tab:id_results_all},  second, third, and fourth columns).  For C/10,  C/6,  and C/3,  the PSO initial guesses are defined by the solution of the identification problem at C/12,  C/10,  and C/6,  respectively.  This methodology allows to identify a sequence of consistent parameters,  avoiding falling in local minima far away from the C/12 solution.  Provided physical consistency of the solutions, C/6 and C/3 scenarios are characterized by faster diffusion and increased overpotentials (Equations \eqref{eq:overp_1} and \eqref{eq:overp_2}) compared to C/12 and C/10.  To incorporate this physical knowledge in the identification problem,  PSO lower bounds for diffusion coefficients and reaction rate constants are modified by using the identified parameters from the previous identification.  Moreover,  contact losses $R_l$ (of the order of few $\mathrm{m\Omega}$ for LFP batteries \cite{prada2012simplified}) are expected to increase as C-rate increases. }

\section{Results}	\label{sec:results}
The model parameters identified for each C-rate of operation are summarized in Table \ref{tab:id_results_all},  {whereas the model parameters that have been provided by the industrial partner of the project (region thicknesses,  maximum solid phase concentrations,  porosities,  active volume fractions,  and transference number) are kept fixed across the various identifications. }

\begin{figure}[!tb]
\centering 
\includegraphics[width = \textwidth]{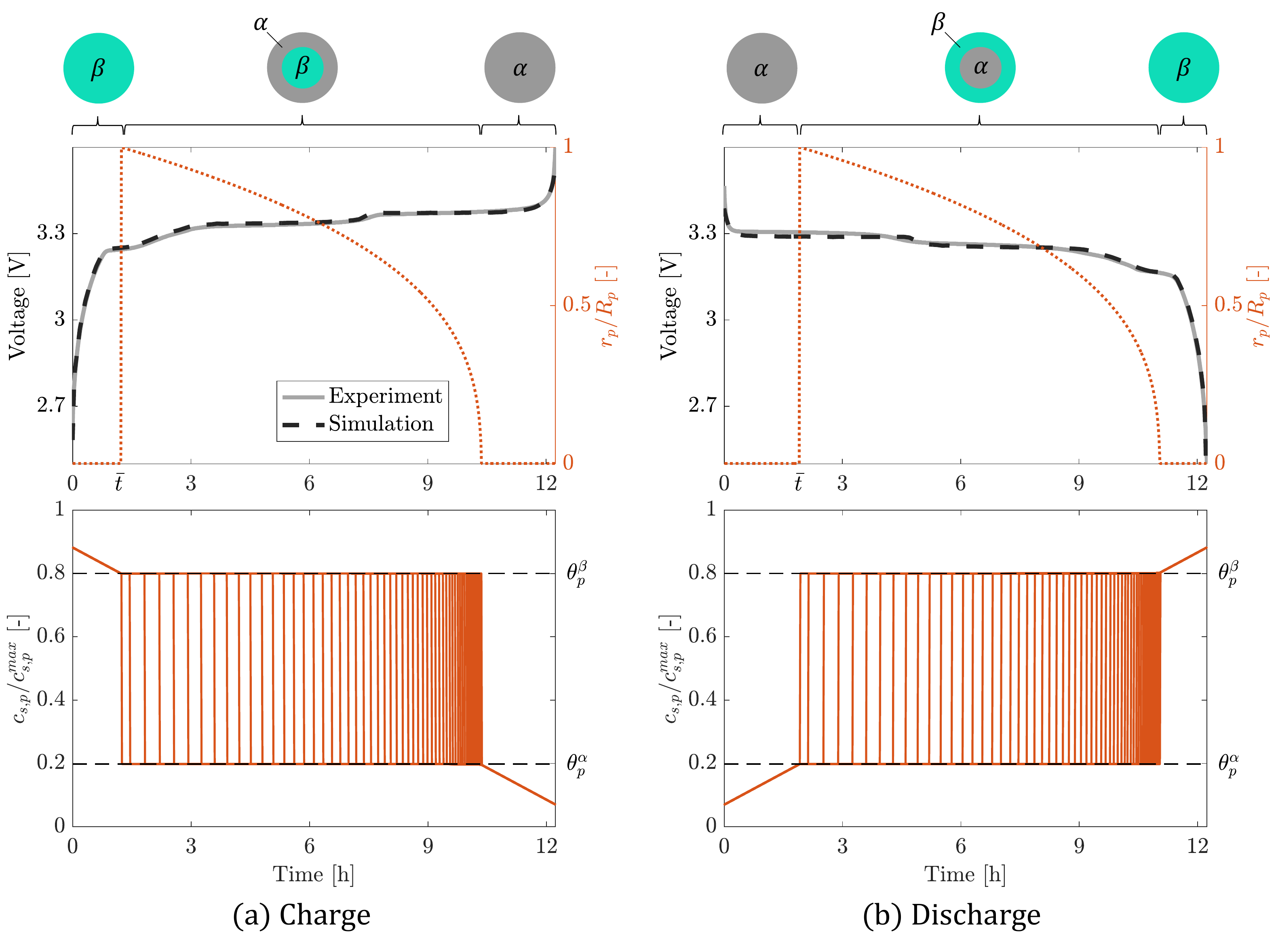}
\caption{Charge and discharge voltage profiles at C/12.  On top,  voltage profiles and moving boundaries are shown.  One-phase regions are associated with $r_p/R_p=0$ and a monotonic decrease (charge) or increase (discharge) of the normalized solid phase concentration $c_{s,p}/c_{s,p}^{max}$.  $r_p/R_p>0$ is associated with the coexistence of two-phase and, therefore, to the core-shrinking process.  In this region, during charge, the concentration moves from $\theta_p^{\beta}$ to $\theta_p^{\alpha}$ (vice versa,  during discharge, $\theta_p^{\alpha}$ transitions to $\theta_p^{\beta}$). For both charge and discharge,  the corresponding phases described in Figures \ref{fig:cs_2} and \ref{fig:cs_1} are highlighted. }
\label{fig:id_results_C12}	
\end{figure}

Identification results for C/12 charge and discharge are shown in Figure \ref{fig:id_results_C12}.  \rev{The solution of the identification problem leads to values for $J_V$ of $0.0021$ and $0.0031$ for charge and discharge, respectively. } Figure \ref{fig:id_results_C12} shows the behavior of the positive electrode solid phase concentration $c_{s,p}$ normalized with respect to $c_{s,p}^{max}$,  and moving boundary $r_p$  normalized with respect to the positive electrode radius $R_p$, where  $r_p/R_p = 0$ corresponds to one-phase regions.  \rev{The solid phase concentration starts  in the one-phase $\beta$ (or $\alpha$) region as the battery starts being charged (or discharged).  As the battery enters the two-phase region,  $r_p/R_p$ becomes greater than zero and the core-shrinking process takes place, i.e., the core shrinks until the one-phase region at the end of charge or discharge is reached. } During charge (discharge),  the core is at uniform concentration $c_{s,p}^\beta$ ($c_{s,p}^\alpha$) and transitions to $c_{s,p}^\alpha$ ($c_{s,p}^\beta$), as the core shrinks.

\begin{figure}[!tb]
\centering 
\includegraphics[width = \textwidth]{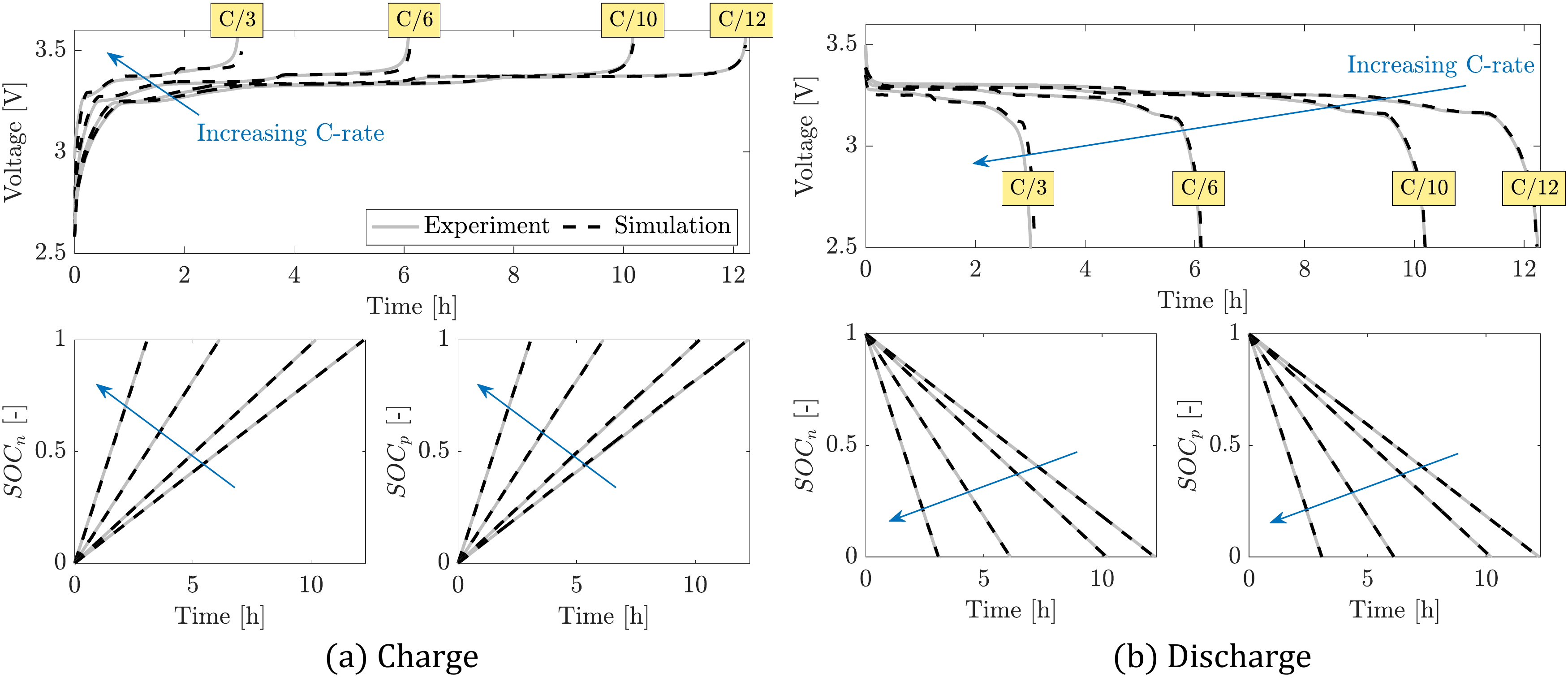}
\caption{Identification results at C/12, C/10, C/6,  and C/3 for charge (a) and discharge (b).}
\label{fig:id_results}	
\end{figure}

\begin{figure}[!tb]
\centering 
\includegraphics[width = \textwidth]{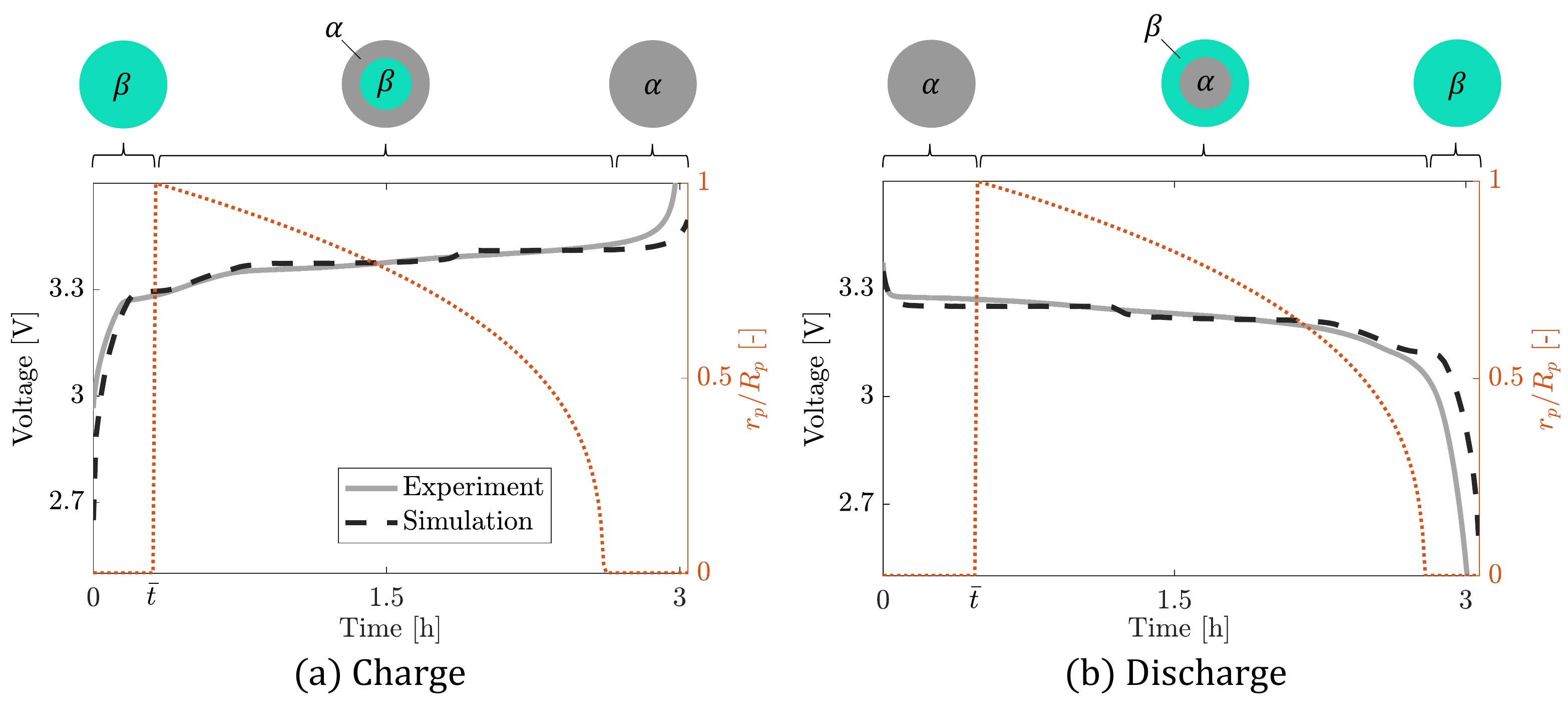}
\caption{Charge and discharge voltage profiles at C/3. The moving boundary is plotted to show the one-phase ($r_p/R_p=0$) and two-phase ($r_p/R_p>0$) regions. } 
\label{fig:id_results_2}	
\end{figure}

In Figure \ref{fig:id_results},  identification results \rev{for C/12,  C/10,  C/6,  and C/3} are compared and corresponding values of the cost function are summarized in Table \ref{tab:id_costs}.   For positive and negative particles,  the model replicates the $SOC$ from Coulomb counting,  with $J_{SOC_n}$ and $J_{SOC_p}$ below 0.005 (or, equivalently, $0.25\ $[Ah]).  \rev{It is noted that for any given C-rate,  $J_{SOC_p}$ is always higher than $J_{SOC_n}$.   This trend is attributed to the presence of the core-shell dynamics,  which makes the numerical solution of the positive particle solid phase concentration and the minimization of the cost term $J_{SOC_p}$ challenging.  Performances of the model with respect to the cell voltage profile are satisfactory for C/12,  C/10,  and C/6,  with the cost function $J_V$ lower than or equal to 0.0054 (or $15\ $[mV]). } As shown by the increasing value of $J$ (last row in Table \ref{tab:id_costs}),  the discrepancy between model outputs and experimental data increases while increasing the C-rate.  \rev{As shown in Figure \ref{fig:id_results_2},  the staging from the graphite OCP becomes more pronounced at C/3,  which determines the discrepancy between experimental and  simulated output voltage profiles. } This prevents the proper convergence of PSO and leads to $J_V$ equal to 0.0215 (or $60\ $[mV]). This behavior is due to limitations of the proposed core-shell ESPM model and will be further analyzed in future works.

\begin{figure}[!tb]
\centering 
\includegraphics[width = \textwidth]{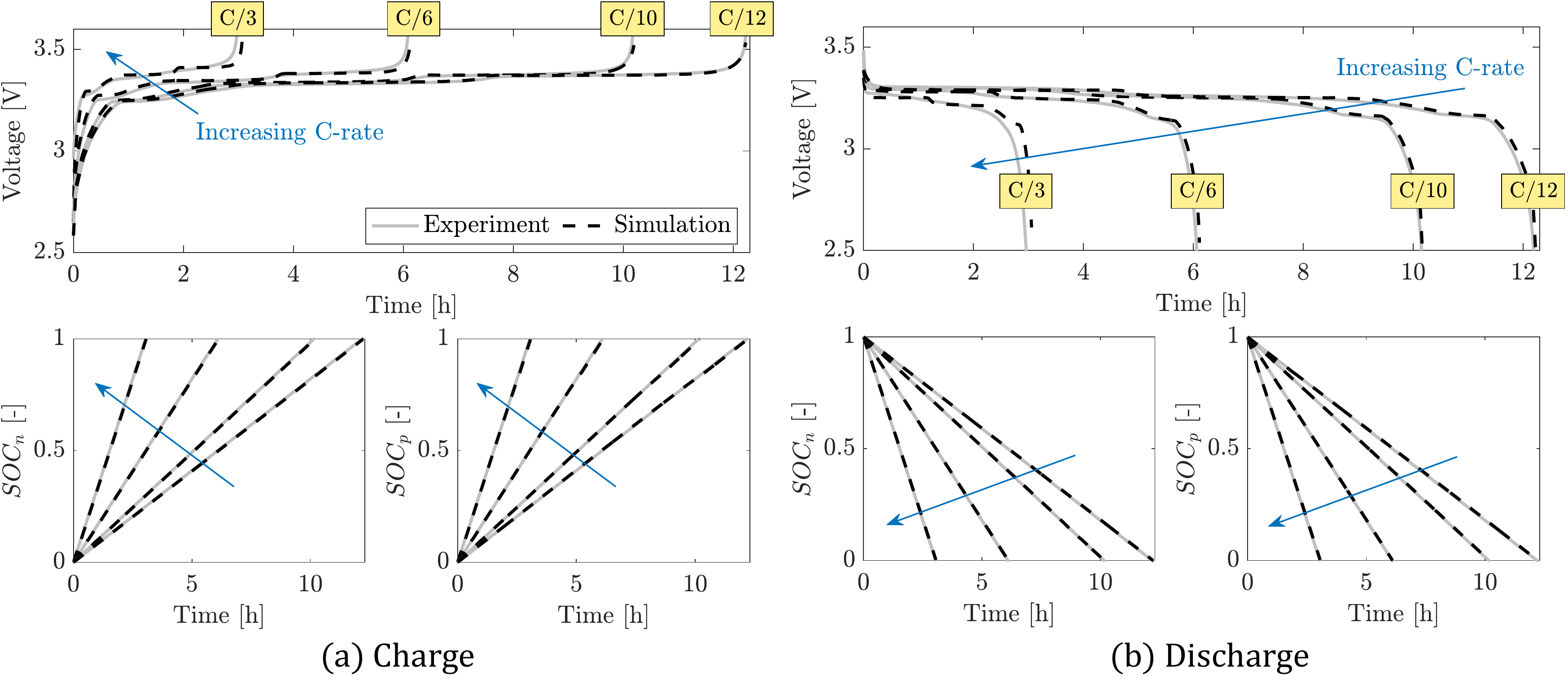}
\caption{Verification results at C/12, C/10, C/6,  and C/3 for charge (a) and discharge (b).}
\label{fig:valid_results}	
\end{figure}

\rev{The model performance is verified against experimental data acquired for a second LFP pouch cell tested at C/12, C/10, C/6, and C/3 constant current.  Results are shown in Figure \ref{fig:valid_results} and corresponding cost function values are summarized in Table \ref{tab:val_costs}. } The performance of the model is satisfactory and trends are consistent with the identification results, i.e., increasing the C-rate the discrepancy between simulated and experimental data increases.  In the worst case,  $J_V$ reaches 0.0294, i.e.,  $80\ $[mV].  

\section{Conclusions}
\rev{This work  presented the development of a core-sell enhanced single particle model for LFP batteries.  Starting from the physical understanding of intercalation and deintercalation phenomena,  governing equations are formulated and discretized using FDM \rev{and FVM}  within a core-shell modeling framework which  assumes a growing shell, or lithium rich phase, and a shrinking core, or lithium poor phase,  during discharge.
The model parameters are identified via a dedicated optimization routine over different C-rate of charge and discharge. }
\rev{Finally, a state-space representation of the core-shell ESPM dynamics is provided,  where it is shown that the positive electrode state vector depends on the moving boundary.  A sensitivity analysis of the solid phase concentration discretization points,  input current sampling time,  and solver tolerances is performed to find a solver setting ensuring stable and robust numerical solutions.  }

The model developed in this paper provides a tool \rev{for the accurate description of LFP phase transition which can be used for state of charge estimation algorithm design.  Future works will be devoted to model hysteresis  and to the validation of the model under real-world driving conditions.}

\section*{Appendix A: relative and absolute tolerance}
\begin{figure}[!tb]
\centering
\includegraphics[width = 1\textwidth]{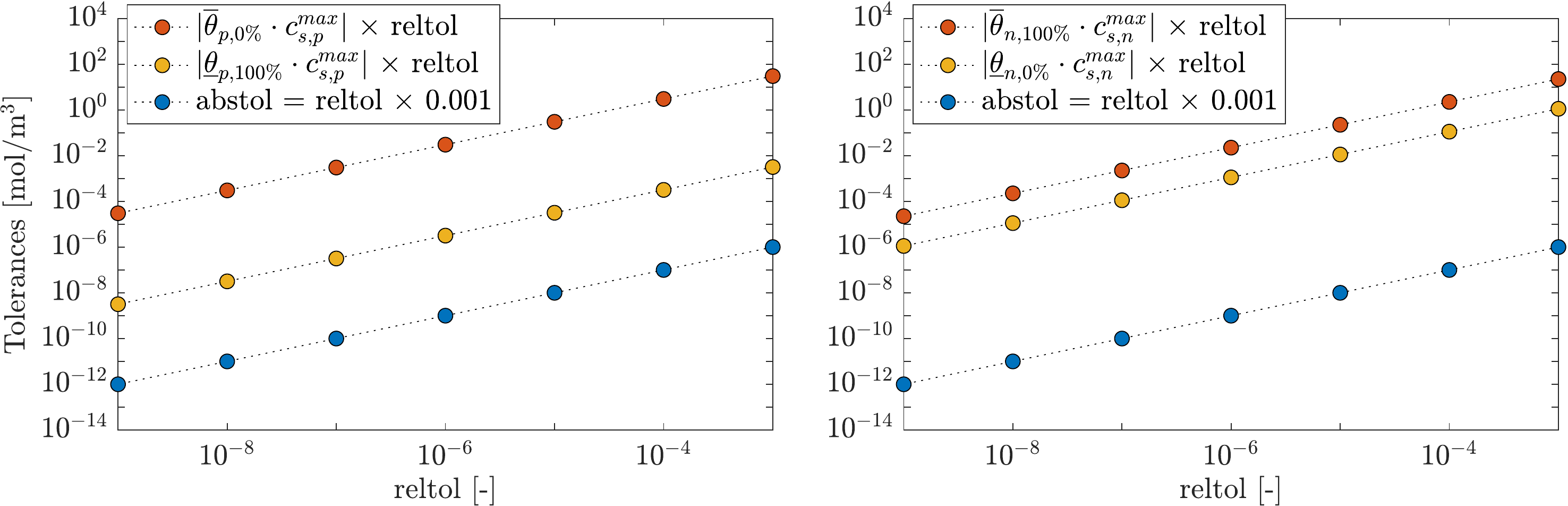}
\caption{Absolute and relative tolerance analysis for positive and negative particle solid phase concentrations.}
\label{fig:cs_tol}
\end{figure}

\begin{figure}[!tb]
\centering 
\includegraphics[width = 1\textwidth]{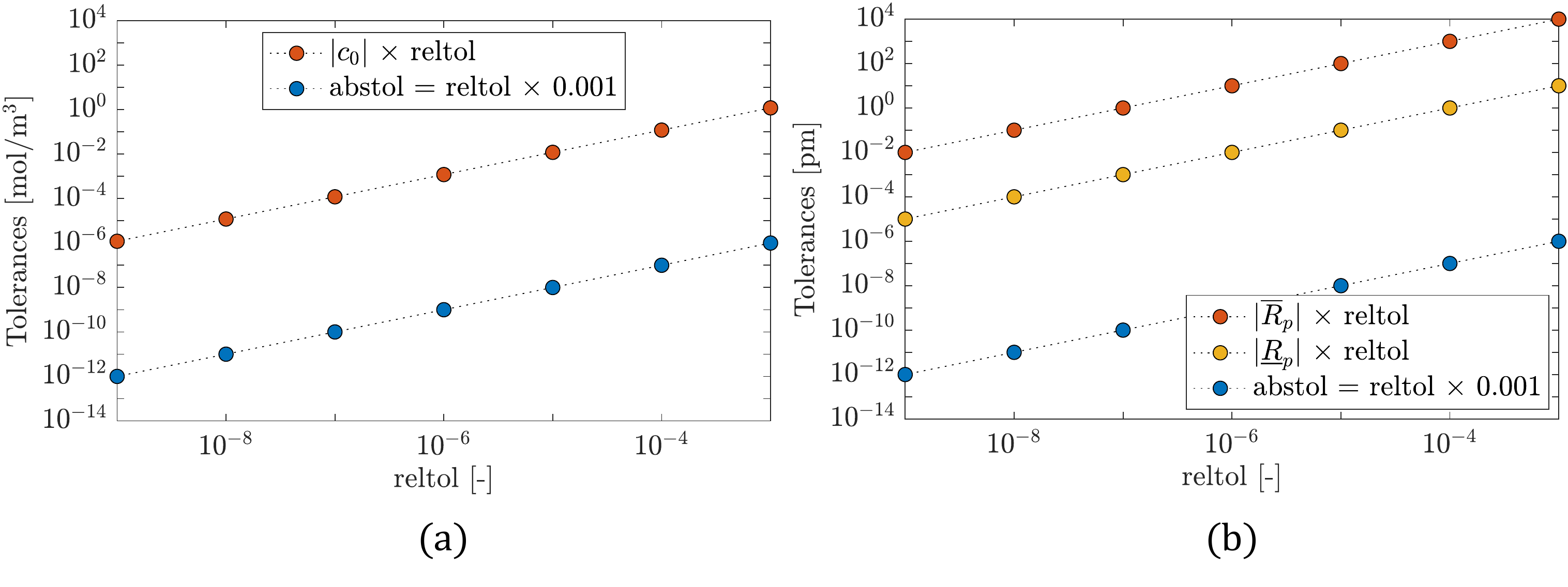}
\caption{Absolute and relative tolerance analysis for (a) electrolyte phase concentration and (b) moving boundary.}
\label{fig:cr_tol}	
\end{figure}

\rev{Relative ($\mathrm{reltol}$) and absolute ($\mathrm{abstol}$) tolerances represent relative and absolute error thresholds used to define the following convergence criterion  for numerical solvers \cite{mathworks2000matlab}:
\begin{equation}
|e| \leq \max{\left(|y|\times\mathrm{reltol},\mathrm{abstol}\right)}
\label{eq:stoppingcrit}
\end{equation}
with $e$ the estimated solver error and $y$ a generic variable.  For practical purposes,  the absolute tolerance should be \myquote{low enough} to be active only when the system is converging.

For the battery electrochemical model developed in this paper,  the numerical convergence of the following variables is analyzed: solid phase concentrations ($\mathbf{c}_{s,p}$ and $\mathbf{c}_{s,n}$),  electrolyte phase concentration ($\mathbf{c}$),  and moving boundary ($r_p$).  For a smooth convergence of the solver the following condition should be satisfied for all the four state variables:
\begin{equation}
|y|\times\mathrm{reltol}>\mathrm{abstol}
\label{eq:convcondition}
\end{equation}
In this appendix,  the analysis is carried out considering the following scaling of the absolute tolerance
\begin{equation}
\mathrm{abstol} = \mathrm{reltol}\times 0.001
\end{equation}
with the relative tolerances in the range $[1\times 10^{-9}, 1\times 10^{-3}]$.  A scaling factor of 0.001 of the $\mathrm{abstol}$ is the default approach used in \textsc{Matlab} \cite{urlmatlab}.  

Starting from the positive particle solid phase concentration ($\mathbf{c}_{s,p}$),  the product $|y|\times\mathrm{reltol}$ in Equation \eqref{eq:convcondition} is computed with respect to the maximum and minimum admissible concentrations:
\begin{equation}
\begin{split}
&\text{Maximum concentration: } |\overline{\theta}_{p,0\%}\cdot c_{s,p}^{max}|\times\mathrm{reltol}\\
&\text{Minimum concentration: }|\underline{\theta}_{p,100\%}\cdot c_{s,p}^{max}|\times\mathrm{reltol}
\end{split}
\label{eq:solvconv_spc_1}
\end{equation} 
where $\overline{\theta}_{p,0\%}$ and $\underline{\theta}_{p,100\%}$ are the upper and lower bounds of the corresponding  stoichiometric coefficients,  as shown in Table \ref{tab:id_results_all}. The same procedure is followed for the negative particle solid phase concentration ($\mathbf{c}_{s,n}$),  with the product $|y|\times\mathrm{reltol}$ defined as:
\begin{equation}
\begin{split}
&\text{Maximum concentration: } |\overline{\theta}_{n,100\%}\cdot c_{s,n}^{max}|\times\mathrm{reltol}\\
&\text{Minimum concentration: }|\underline{\theta}_{n,0\%}\cdot c_{s,n}^{max}|\times\mathrm{reltol}
\end{split}
\label{eq:solvconv_spc_2}
\end{equation}
where $\overline{\theta}_{n,100\%}$ and $\underline{\theta}_{n,0\%}$ are the upper and lower bounds of the corresponding  stoichiometric coefficients,  as shown in Table \ref{tab:id_results_all}.  Results for Equations \eqref{eq:solvconv_spc_1} and \eqref{eq:solvconv_spc_2} as a function of $\mathrm{reltol}$ are shown in Figure \ref{fig:cs_tol}.  A relative tolerance higher than $1\times 10^{-5}$ leads to values of \eqref{eq:solvconv_spc_1} and \eqref{eq:solvconv_spc_2} higher than $1~[\mathrm{mol/m^3}]$ (or $6.94~[\mathrm{g/m^3}]$,  given the molar mass of lithium equal to $6.94~[\mathrm{g/mol}]$).  Hence,  to ensure high accuracy solutions,  the following constraint is enforced:
\begin{equation}
\mathrm{reltol} \leq 1\times 10^{-5}
\label{eq:constrreltol}
\end{equation} 
which is used in Tables \ref{tab:sens1} and \ref{tab:sens2},  to define the relative tolerance upper bound.

Figure \ref{fig:cr_tol} shows the product $|y|\times\mathrm{reltol}$ for the electrolyte phase concentration $\mathbf{c}$ (a) and the moving boundary $r_p$ (b).  For the electrolyte phase concentration,  $|c_0|\times\mathrm{reltol}$,  {with $c_0 = 1200\ [\mathrm{mol/m^3}]$ the initial electrolyte concentration from \cite{prada2013simplified}},  is computed and compared to the absolute tolerance.   
For the moving boundary,  $|\overline{R}_p|\times\mathrm{reltol}$ and $|\underline{R}_p|\times\mathrm{reltol}$ are calculated,  with $\overline{R}_p$ and $\underline{R}_p$ the positive particle radius upper and lower bounds defined in Table \ref{tab:id_results_all}\footnote{$\overline{R}_p$ and $\underline{R}_p$ are scaled from \textit{meter} to \textit{picometer}.}. 

Ultimately,  constraint \eqref{eq:convcondition} is satisfied for solid phase concentrations,  electrolyte phase concentration,  and moving boundary. }

\section*{Appendix B: sensitivity analysis}
\rev{The sensitivity analysis is divided in two steps.  Step 1 is used to select values for sampling time ($dt$) and absolute tolerance ($\mathrm{abstol}$).  Step 2 is used to select the pair $\{N_r,\mathrm{reltol}\}$ ensuring a robust solution of the model equations.  For each solver setting $\mathcal{C}$,  the ODEs solution accuracy is defined with respect to the objective function in Equation \eqref{eq:costterms},  for the C/12 constant current discharge profile.}

\noindent\textbf{Step 1: Selection of $\mathbf{abstol}$ and $\boldsymbol{dt}$.  }
\rev{The sensitivity is performed with respect to all combinations of solver settings in Table \ref{tab:sens1},  considering the initial guess $\Theta_1$ for the parameter vector (Table \ref{tab:id_results_all}, column four).  }

\begin{figure}[!tb]
\centering 
\includegraphics[width = 1\textwidth]{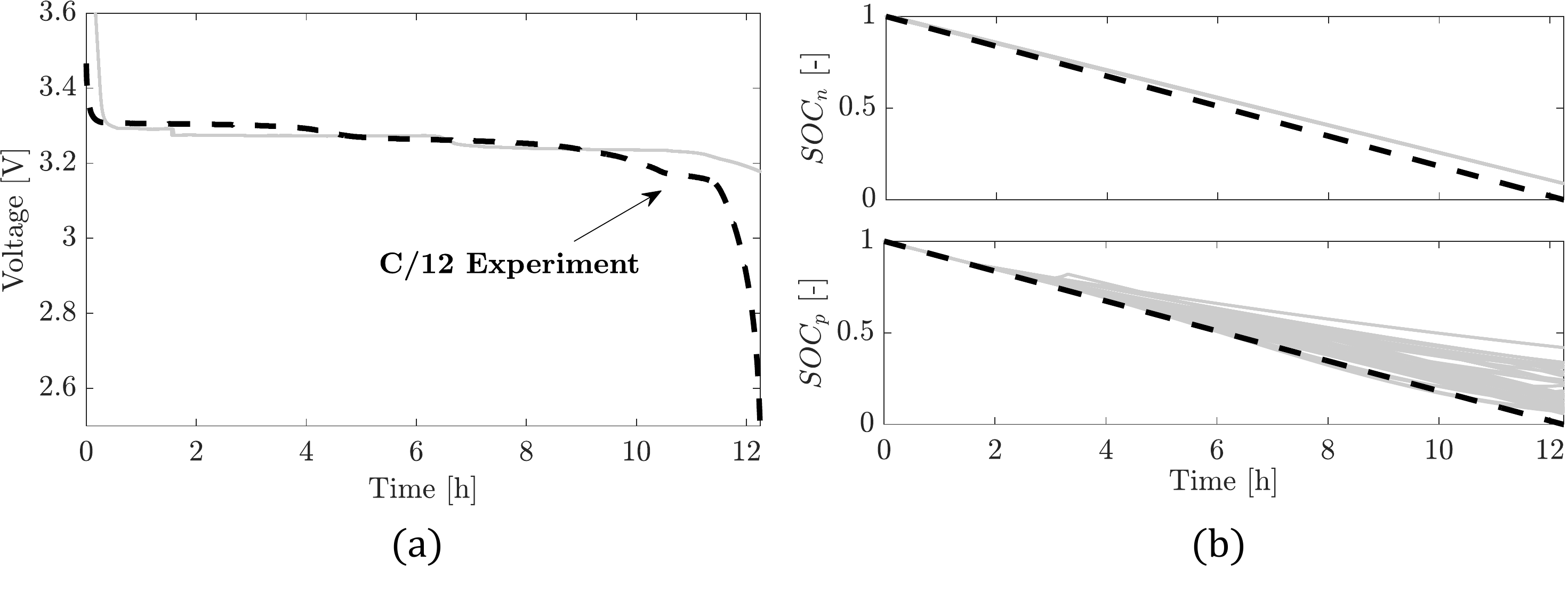}
\caption{Simulated open circuit voltage (a) and $SOC$ (b) for positive and negative electrodes.  \rev{Black-dashed lines show the experimental voltage and $SOC$ from Coulomb counting at C/12 ($\mathrm{abstol}=\mathrm{reltol}\times 0.001$).}}
\label{fig:sim_vsoc}	
\end{figure}

\rev{For each setting $\mathcal{C}$ in Table \ref{tab:sens1},  the simulated voltage and $SOC$ for positive and negative particles are shown in Figure \ref{fig:sim_vsoc}	 and compared to C/12 experimental voltage and $SOC$ from Coulomb counting (black-dashed lines).  For some combinations $\mathcal{C}$,  the solution is unstable and a nonlinear-diverging $SOC$ profile is obtained (Figure \ref{fig:sim_vsoc}(b)). } This behavior, visible in the positive particle only,  proves the challenges introduced by the core-shell dynamics and the need to perform the sensitivity analysis. 

\begin{figure}[!tb]
\centering 
\includegraphics[width = 1\textwidth]{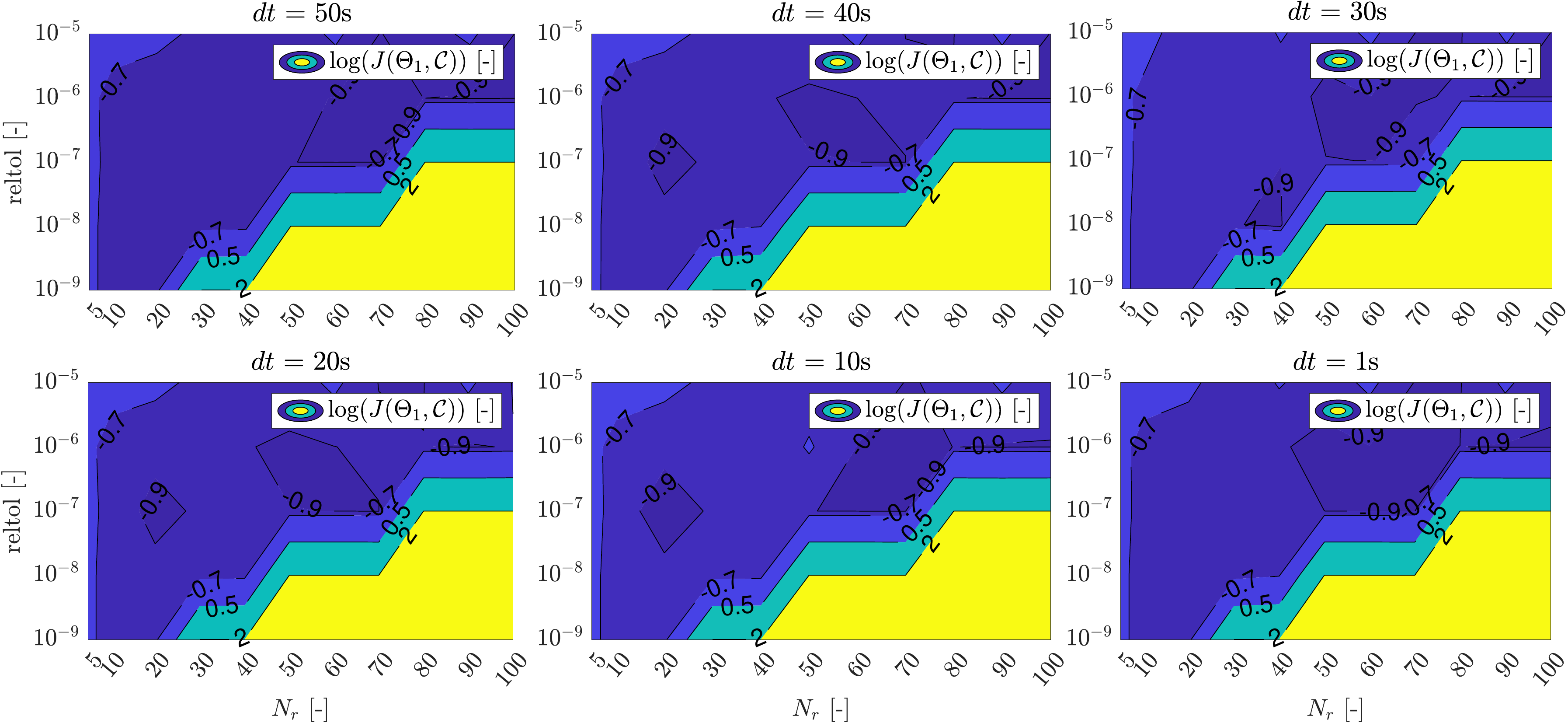}
\caption{Logarithm of the cost function $J(\Theta_1,\mathcal{C})$ as a function of $N_r$ and $\mathrm{reltol}$ ($\mathrm{abstol=reltol}\times 0.001)$.}
\label{fig:Jsens}	
\end{figure}

\begin{figure}[!tb]
\centering 
\includegraphics[width = 0.5\textwidth]{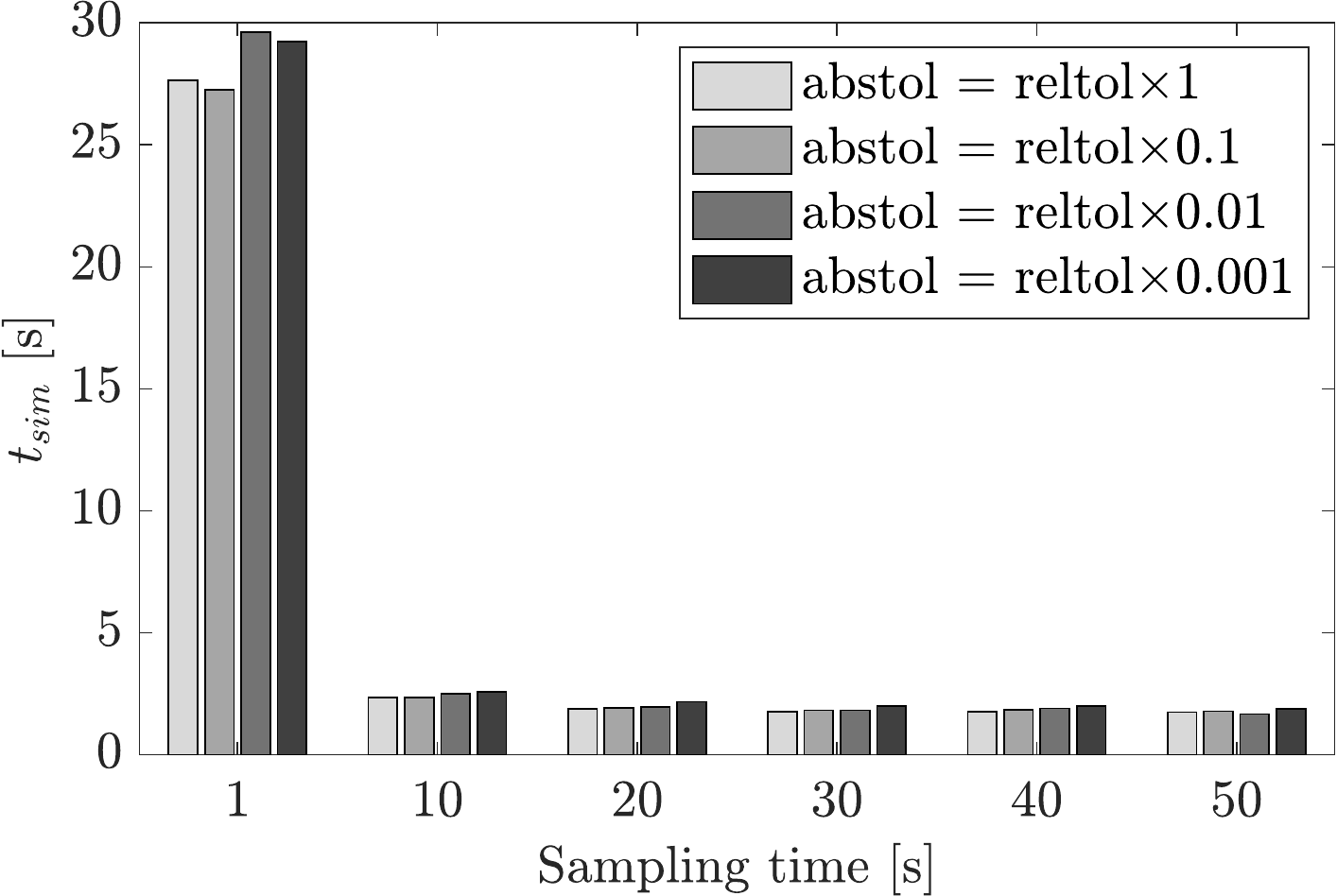}
\caption{Average simulation time $t_{sim}$ as a function of the sampling time $dt$, for each absolute tolerance scaling.}
\label{fig:tsimsens}	
\end{figure}

\rev{Figure \ref{fig:Jsens} shows the cost function $J(\Theta_1,\mathcal{C})$ for $\mathrm{abstol=reltol}\times 0.001$,  considering all the combinations in Table \ref{tab:sens1}.   Each contour plot is associated with a different current profile sampling time $dt$. } The yellow region highlights where the solver is not converging. To this region,  we enforce a fictitious high cost equal to $J(\Theta_1,\mathcal{C})=100$.  Lack of convergence is shown for high $N_r$ (which leads to high order ODEs systems) and low relative tolerances (requiring higher solution accuracy before the stopping criterion \eqref{eq:stoppingcrit} is met).   \rev{As shown in Figure \ref{fig:Jsens}},  a sampling time increase does not change the convergence region of the solver.  However,  a reduction of one order of magnitude of the sampling time (e.g., from 10 to 1$\ $[s]) leads to a one order of magnitude increment of the simulation time (from 3 to 30$\ $[s]).  This is shown in Figure \ref{fig:tsimsens},  where,  for each sampling time $dt$ and various absolute tolerance scaling factors,  the average simulation time ($t_{sim}$) is computed over the $N_r$-$\mathrm{reltol}$ region in which the solver is converging (i.e.,  blue regions in Figure \ref{fig:Jsens}).

\begin{figure}[!tb]
\centering 
\includegraphics[width = 0.8\textwidth]{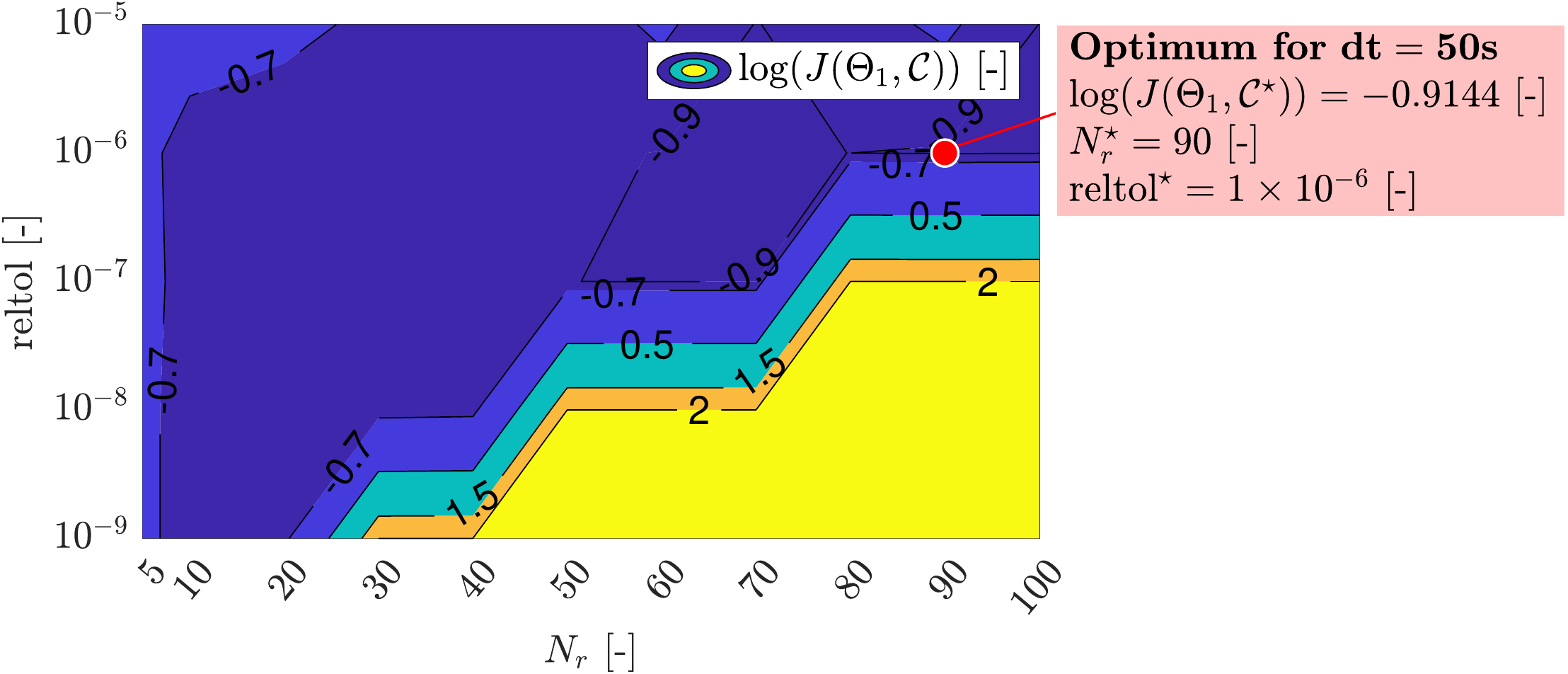}
\caption{Optimal $\{N_r^\star,\mathrm{reltol}^\star\}_1$  for $dt=50\mathrm{s}$ ($\mathrm{abstol=reltol}\times 0.001$).}
\label{fig:optsol}	
\end{figure}

\begin{figure}[!tb]
\centering 
\includegraphics[width = 0.8\textwidth]{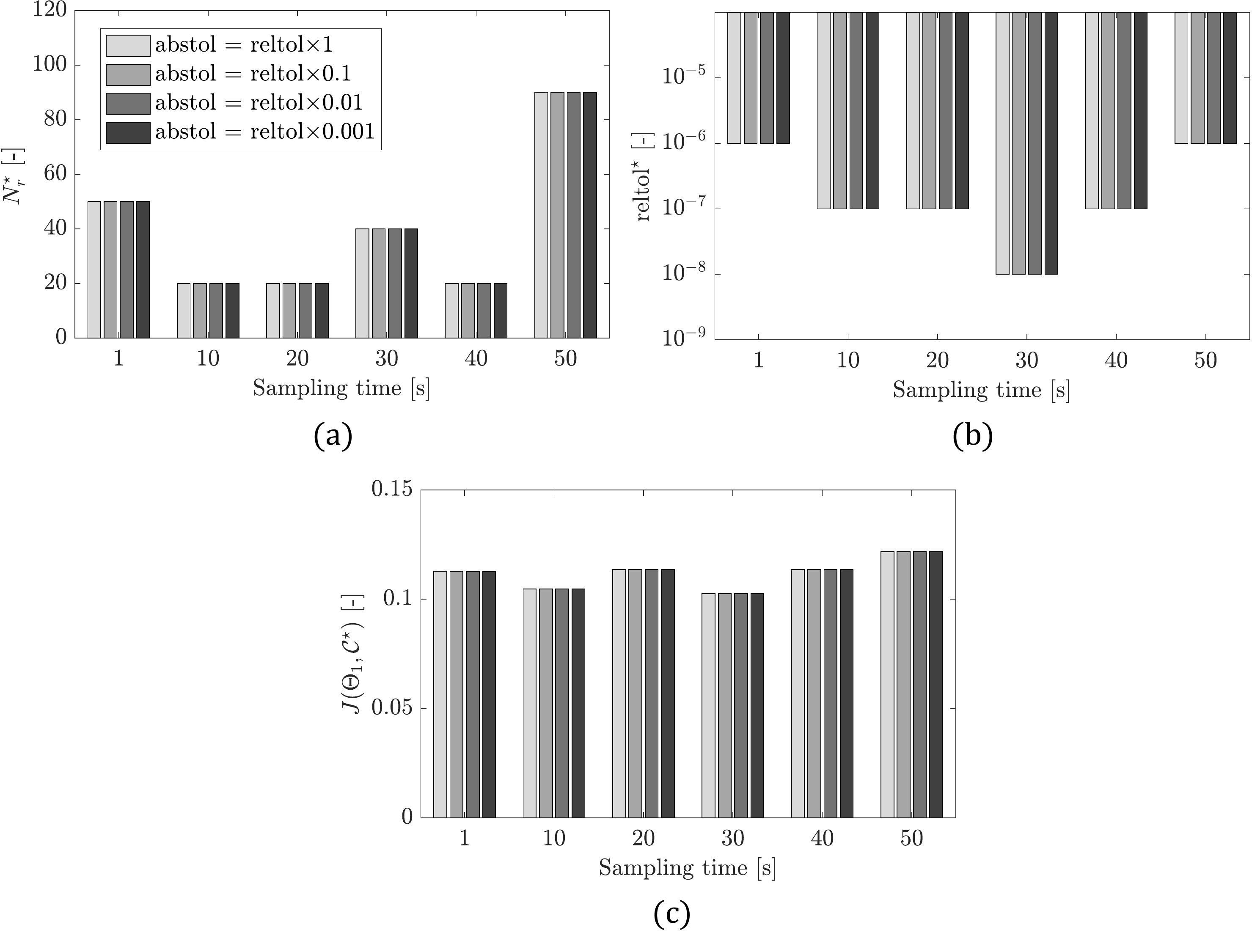}
\caption{Optimal $\{N_r^\star,\mathrm{reltol}^\star\}_1$ and corresponding $J(\Theta_1,C^\star)$ as a function of the sampling time $dt$, for each absolute tolerance scaling.}
\label{fig:nopert_alltogether}	
\end{figure}

For a given sampling time and absolute tolerance scaling factor,  the optimal $\{N_r^\star,\mathrm{reltol}^\star\}_1$ setting is the one minimizing the cost function $J(\Theta_1,\mathcal{C})$:
\begin{equation}
\underset{\{N_r,\mathrm{reltol}\}}{\mathbf{minimize}}\ J(\Theta_1,\mathcal{C})
\label{eq:J1}
\end{equation}
In Figure \ref{fig:optsol}, the optimal setting for $dt=50\ $[s] and $\mathrm{abstol} = \mathrm{reltol}\times 0.001$ is shown.  In this scenario,  the cost function is minimized for:
\rev{\begin{equation}
N_r^\star = 90, \ \mathrm{reltol}^\star = 1\times 10^{-6}\
\end{equation}
with log($J(\Theta_1,\mathcal{C}^\star)\text{)} = -0.9144 \rightarrow\ J(\Theta_1,\mathcal{C}^\star) = 0.1218$.  Results from repeating the analysis for each combination of $dt$ and $\mathrm{abstol}$ are shown in Figure \ref{fig:nopert_alltogether}.  Different scaling factors of the absolute tolerance do not affect $N_r^\star$ and $\mathrm{reltol}^\star$,  also, the cost function value at the optimum point  is not changing macroscopically while varying the sampling time.  Hence,  the following choices are made:
\begin{equation}
dt = 50\ [\mathrm{s}],\  \mathrm{abstol} = \mathrm{reltol}\times 0.001
\label{eq:bestfromstep1}
\end{equation}}

\noindent\textbf{Step 2: Selection of $\mathbf{reltol}$ and $\boldsymbol{N_r}$.  }
\rev{The same analysis proposed in Step 1 is repeated for 600 realizations $\Theta_\iota$ of the model parameter vector (with $\iota=1,...,600$).  Each realization $\Theta_\iota$ takes the following expression:
\begin{equation}
\Theta_{\iota} = \big[ R_n, R_p,A_{cell},D_{s,n},D_{s,p}, \theta_{n,100\%},\theta_{n,0\%},\theta_{p,100\%},\theta_{p,0\%},\theta_p^{\alpha},\theta_p^{\beta},R_l\big]^T
\label{eq:iota}
\end{equation}
For each $\Theta_\iota$,  parameters are randomly picked inside bounds defined in Table \ref{tab:id_results_all} and all combinations of solver settings in Table \ref{tab:sens2} are tested.  As a result of Step 1,  absolute tolerance scaling and sampling time are fixed (Equation \eqref{eq:bestfromstep1}).}

Similarly to Equation \eqref{eq:J1}, for each realization $\Theta_\iota$, the optimal pair $\{N_r^\star,\mathrm{reltol}^\star\}_\iota$ is the one minimizing the cost function $J(\Theta_\iota,\mathcal{C})$:
\begin{equation}
\underset{\{N_r,\mathrm{reltol}\}}{\mathbf{minimize}}\ J(\Theta_\iota,\mathcal{C})
\label{eq:J2}
\end{equation}
The optimum of problem \eqref{eq:J2} is a function of $\iota$ and,  for each vector $\Theta_\iota$,  a different optimal pair $\{N_r^\star,\mathrm{reltol}^\star\}_\iota$ is found.  

\rev{Our goal is to define a solver setting $\hat{\mathcal{C}}$ suitable for most scenarios, i.e., parameter vectors $\Theta_\iota$.  To this aim,}  we rewrite the problem in a probabilistic framework and, upon a frequentist approach, we define the \myquote{Probability of $\{N_r^\star,\mathrm{reltol}^\star\}$}  as the probability that a certain configuration $\{N_r,\mathrm{reltol}\}$ is optimal.  

The probability is defined as $\mathcal{P}(\{N_r^\star,\mathrm{reltol}^\star\})$ and takes the following expression:
\begin{equation}
\mathcal{P}(\{N_r^\star,\mathrm{reltol}^\star\}) = \frac{\mathcal{N}_{\{N_r^\star,\mathrm{reltol}^\star\}}}{\mathcal{N}}\cdot 100
\end{equation}
where $\mathcal{N}_{\{N_r^\star,\mathrm{reltol}^\star\}}$ is the number of times,  given the set of tested vectors $\Theta_\iota$,  a solver setting $\{N_r^\star,\mathrm{reltol}^\star\}$ is optimal and $\mathcal{N}$ is the total number of realizations $\Theta_\iota$ considered in the analysis, i.e., 600.

\begin{figure}[!tb]
\centering 
\includegraphics[width = 0.8\textwidth]{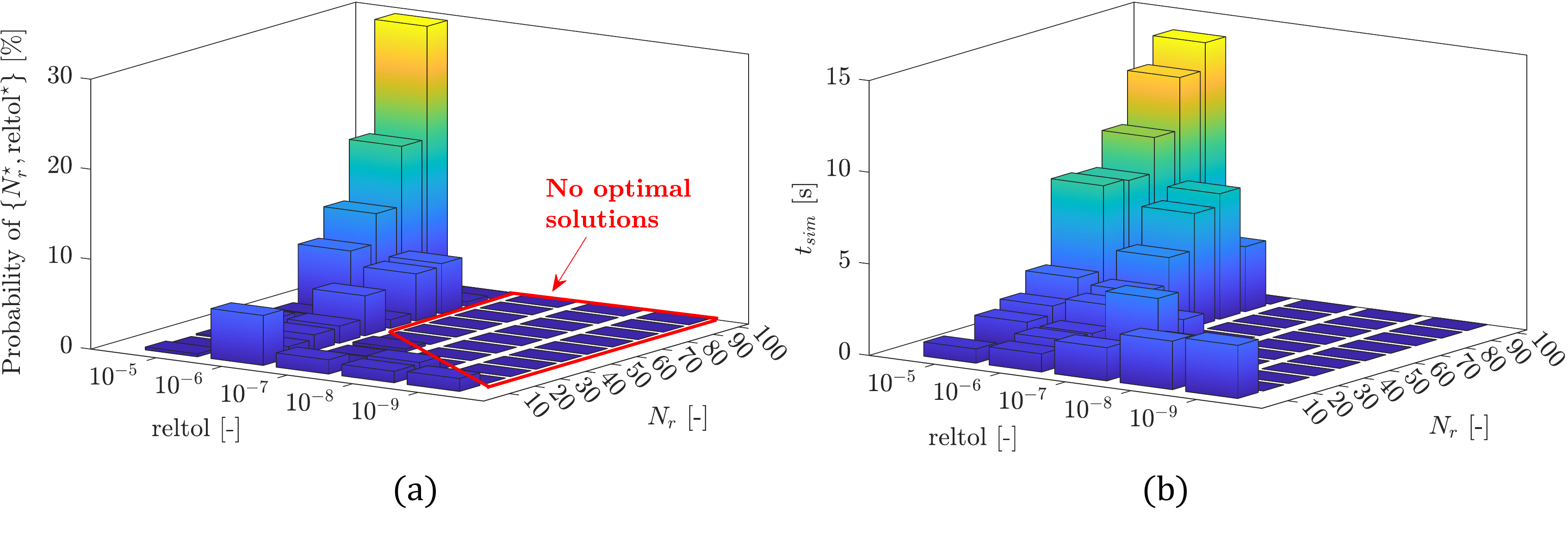}
\caption{Probability of $\{N_r^\star,\mathrm{reltol}^\star\}$ to be optimal (a) and average simulation time $t_{sim}$ (b).  The dark blue region inside the red polygon in (a) highlights the settings that are never optimal due to lack of convergence of the solver. }
\label{fig:pert_alltogether}	
\end{figure}

The probability distribution is shown in Figure \ref{fig:pert_alltogether}(a),  together with the  average simulation time $t_{sim}$ in Figure \ref{fig:pert_alltogether}(b).  Most of the optimal solutions are in the high relative tolerance region,  between $1\times 10^{-6}$ and $1\times 10^{-5}$.  This is reasonable since,  for these tolerances,  the problem becomes easier to solve and convergence of the solver is more likely.  \rev{The dark blue region inside the red polygon in Figure \ref{fig:pert_alltogether}(a)  highlights the solver settings for which there are no optimal solutions due to lack of convergence of the solver.  This happens for high values of $N_r$ and low relative tolerances.} From Figure  \ref{fig:pert_alltogether}(b),  it is seen that increasing the number of discretization points $N_r$ leads to an increase of the computational time as a higher order ODEs system is to be solved.  

\begin{figure}[!tb]
\centering   
\includegraphics[width = 0.5\textwidth]{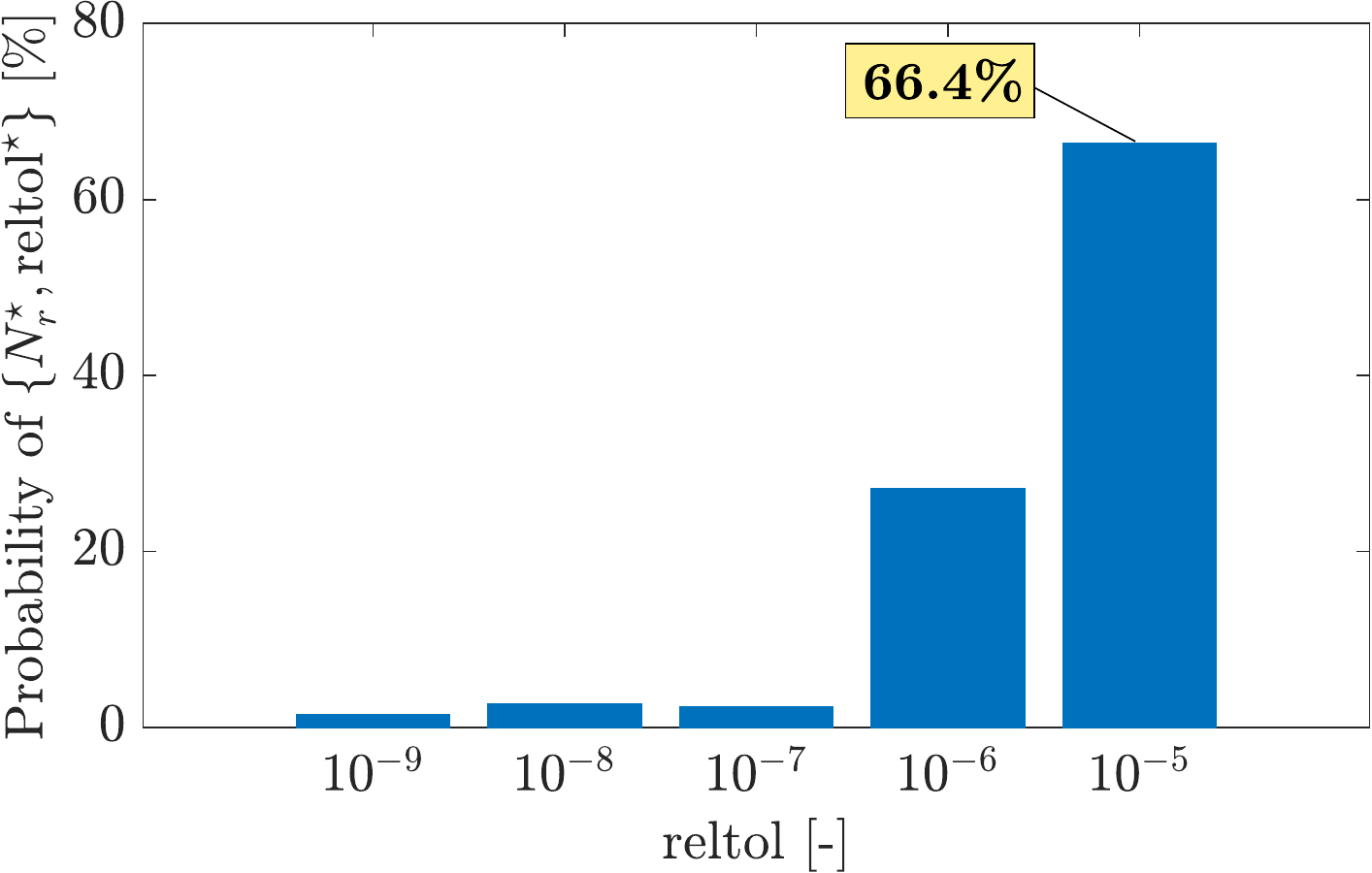}
\caption{Cumulative probabilities grouped by relative tolerance.}
\label{fig:pert_barprob}	
\end{figure}

\begin{figure}[!tb]
\centering 
\includegraphics[width = 0.5\textwidth]{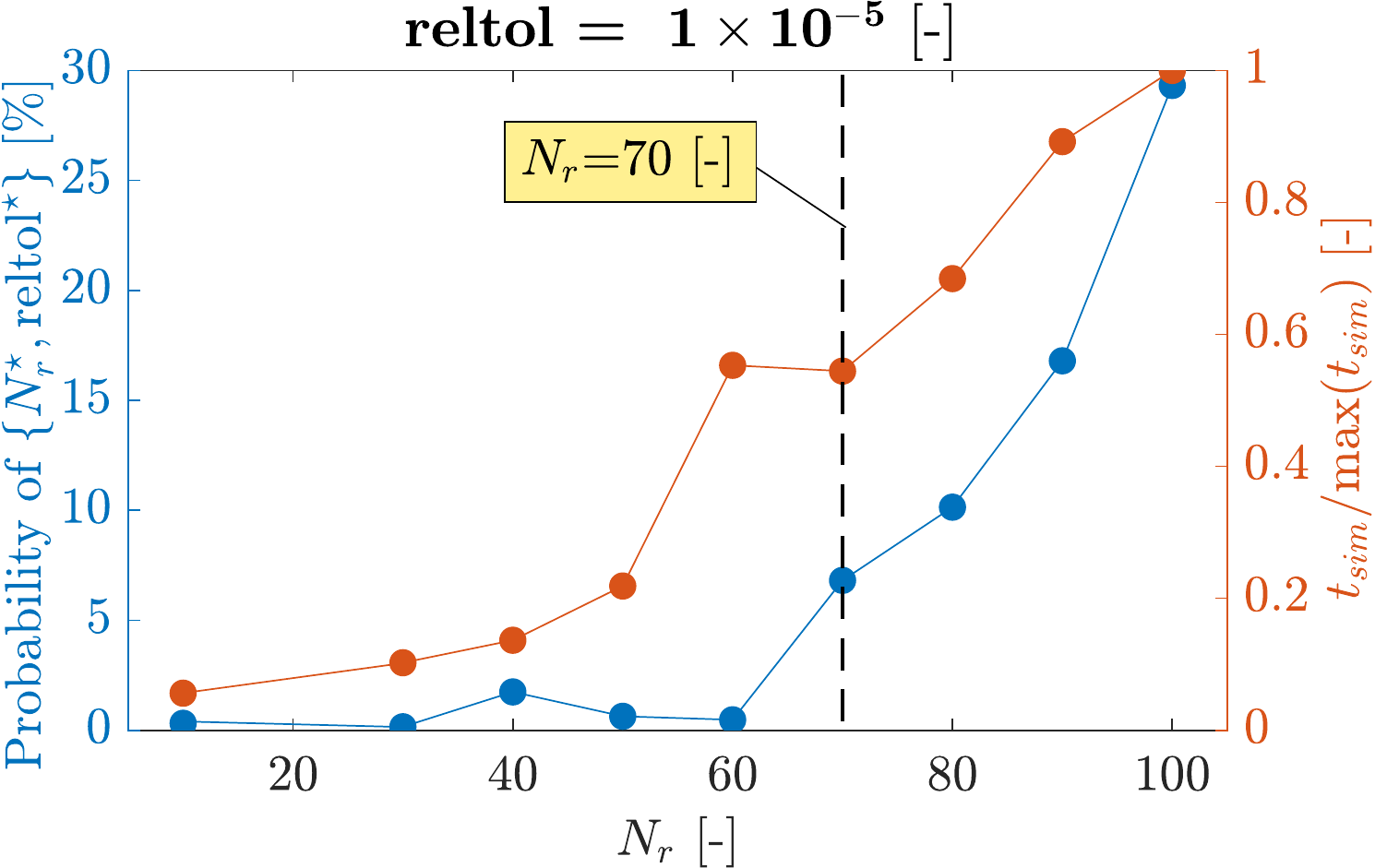}
\caption{Probability of $\{N_r^\star,\mathrm{reltol}^\star\}$ and normalized average simulation time ($\mathrm{reltol} = 1\times 10^{-5}$ [-]).  The trade-off $N_r = 70$ is highlighted.}
\label{fig:pert_tradeoff}	
\end{figure}

\rev{Summing the probabilities in Figure \ref{fig:pert_alltogether}(a) along $N_r$,  while grouping by the relative tolerance,  leads to the cumulative probabilities in Figure \ref{fig:pert_barprob}, from which it is found that $66.4\%$ of the optimal solutions are at $\mathrm{reltol}=1\times 10^{-5}$,  which is chosen as the best relative tolerance setting.  According to Figure \ref{fig:pert_tradeoff},  $N_r = 70$  is associated with the best trade-off between probability and computational time -- normalized with respect to the maximum simulation time at $N_r = 100$ -- and is chosen as the best setting for the radial coordinate discretization points.}

\rev{Ultimately,  Steps 1 and 2 define the best solver setting $\hat{\mathcal{C}}$ for the ODEs system solution:}
\begin{equation}
\hat{\mathcal{C}} = \{70,50,1\times 10^{-5},\mathrm{reltol}\times 0.001\}
\end{equation}

\section*{Appendix C: identifiability of model parameters}
\begin{figure}[!tb]
\centering 
\includegraphics[width = 0.5\textwidth]{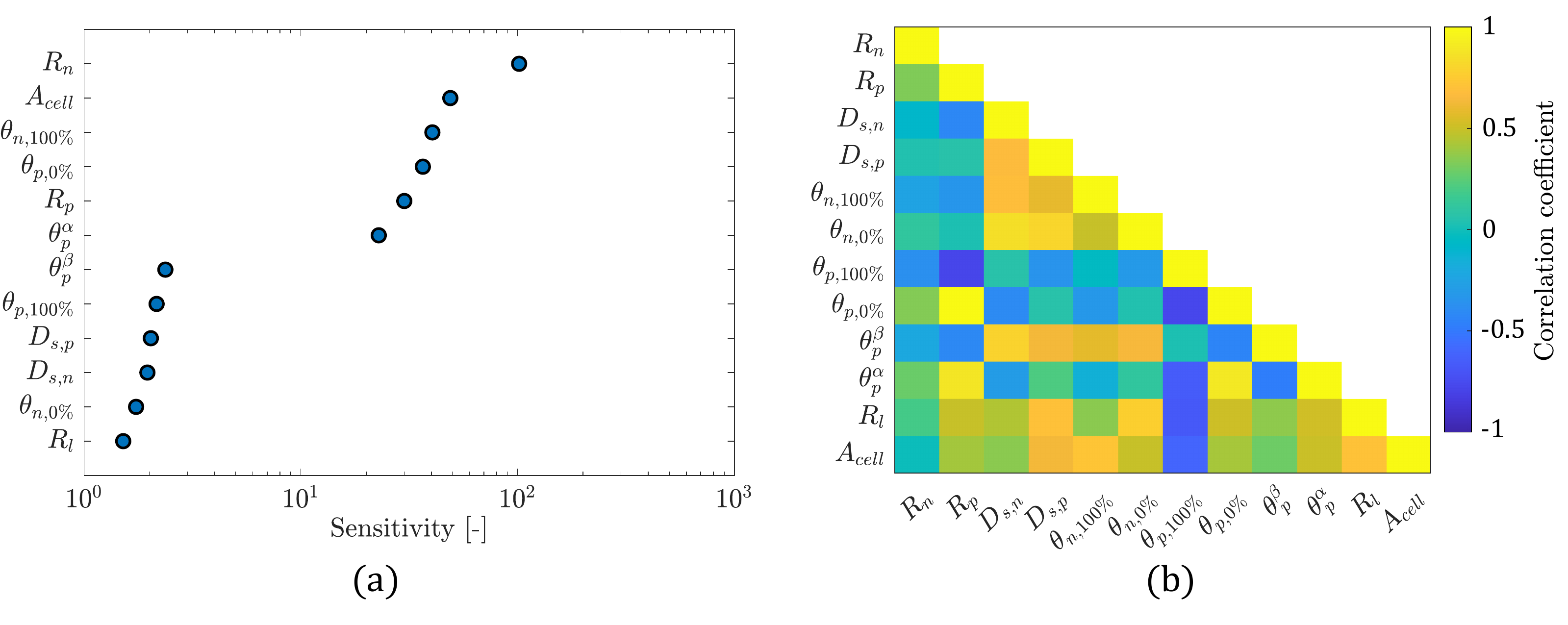}
\caption{\rev{Sensitivity of the model output with respect to small perturbations of the C/12 identified parameters (a) and correlation analysis of the core-shell ESPM parameters (b).}}
\label{fig:senspar}	
\end{figure}

\rev{To assess the sensitivity of the model output with respect to small perturbations of the identified parameters,  a sensitivity analysis is carried out considering the C/12 constant current scenario.  Identified parameters in Table \ref{tab:id_results_all} (column five) are used as nominal condition and perturbed to compute the sensitivity matrix.   The sensitivity of each parameter,  computed as the Euclidean norm along the columns of the sensitivity matrix,  is shown in Figure \ref{fig:senspar}(a).  Following the rationale of \cite{allam2020online},  starting from the sensitivity matrix,  a correlation analysis is performed.  Figure \ref{fig:senspar}(b) allows to understand if perturbations of different parameters result in the same output response.  If this happens,  correlation coefficients assume high values ($>  0.8$ or $< -0.8$),  meaning that parameters can not be identified uniquely.  Generally,  this correlation analysis is used to reduce the vector of identified parameters.  However,  the cell under investigation is still in the design phase and most of its parameters are unknown,  hence,  results in Figure \ref{fig:senspar} should be interpreted just as indicators of the goodness of the identification.  Moreover,  the correlation analysis is performed \textit{a posteriori} and is a function of the nominal parameter vector and perturbation strategy.  The analysis of other methodologies for practical and structural identifiability  is left to future works \cite{miao2011identifiability}.}

\bibliographystyle{unsrt}

\section*{Nomenclature}
\entry{$\mathrm{a,b}$}{Intercalation/deintercalation reactions' coefficients, $[\mathrm{-}]$}
\entry{$brugg$}{Bruggeman coefficient, $[\mathrm{-}]$}
\entry{$N$}{Number of samples, $[\mathrm{-}]$}
{\entry{$N_{x,i}$}{Number of discretization points along $x$ ($i \in{\mathcal{M}}$),  $[\mathrm{-}]$}}
\entry{$N_{r,i}$}{Number of discretization points along $r$ ($i \in{\hat{\mathcal{M}}}$),  $[\mathrm{-}]$}
\entry{$\mathcal{N}$}{Counter,  $[\mathrm{-}]$}
\entry{$\mathcal{P}$}{Probability,  $[\mathrm{-}]$}
\entry{$SOC_i$}{State of charge ($i \in\hat{\mathcal{M}}$), $[\mathrm{-}]$}
\entry{$t_+$}{Transference number, $[\mathrm{-}]$}
\entry{$w_1,w_2,w_3$}{Cost function weights,  $[\mathrm{-}]$}
\entry{$\varepsilon_i$}{Porosity ($i \in{\mathcal{M}}$), $[\mathrm{-}]$}
\entry{$\nu_{i}$}{Solid phase active volume fraction ($i \in\hat{\mathcal{M}}$), $[\mathrm{-}]$}
\entry{$\nu_{i,filler}$}{Filler active volume fraction ($i \in\hat{\mathcal{M}}$), $[\mathrm{-}]$}
\entry{$\theta_i^{surf}$}{Surface normalized lithium concentration ($i \in\hat{\mathcal{M}}$), $[\mathrm{-}]$}
\entry{$\theta_i^{bulk}$}{Bulk normalized lithium concentration ($i \in\hat{\mathcal{M}}$), $[\mathrm{-}]$}
\entry{$\theta_{i,0\%}$}{Reference stoichiometry ratio at 0\% $SOC$ ($i \in\hat{\mathcal{M}}$), $[\mathrm{-}]$}
\entry{$\theta_{i,100\%}$}{Reference stoichiometry ratio at 100\% $SOC$ ($i \in\hat{\mathcal{M}}$), $[\mathrm{-}]$}
\entry{$\theta_{p}^\alpha,\theta_{p}^\beta$}{Positive particle $\alpha$ and $\beta$ phase normalized concentrations, $[\mathrm{-}]$}
\entry{$\chi$}{Normalized coordinate, $[\mathrm{-}]$}
\entry{$t,dt$}{Time and its differential, $[\mathrm{s}]$}
\entry{$\bar{t}$}{Time instant of the transition from one-phase to two-phase,  $[\mathrm{s}]$}
\entry{$\epsilon$}{Small enough $R_p$ correction \cite{takahashisub}, $[\mathrm{m}]$}
\entry{$x$}{Cartesian coordinate, $[\mathrm{m}]$}
\entry{$r$}{Radial coordinate, $[\mathrm{m}]$}
\entry{$r_p$}{Moving boundary, $[\mathrm{m}]$}
\entry{$L_i$}{Region thickness ($i \in\mathcal{M}$), $[\mathrm{m}]$}
\entry{$R_i$}{Particle radius ($i \in\hat{\mathcal{M}}$), $[\mathrm{m}]$}
\entry{$A_{cell}$}{Cell cross section area, $[\mathrm{m}^2]$}
\entry{$a_{i}$}{Specific surface area ($i \in\hat{\mathcal{M}}$), $[\mathrm{m^2/m^3}]$}
\entry{$D$}{Electrolyte phase diffusion coefficient, $[\mathrm{m^2/s}]$}
\entry{$D_i^{eff}$}{Effective electrolyte phase diffusion coefficient ($i \in\mathcal{M}$), $[\mathrm{m^2/s}]$}
\entry{$D_{s,i}$}{Solid phase diffusion coefficient  ($i \in\hat{\mathcal{M}}$), $[\mathrm{m^2/s}]$}
\entry{$c$}{Electrolyte concentration, $[\mathrm{mol/m^3}]$}
\entry{$c_{s,i}$}{Solid phase concentration ($i \in\hat{\mathcal{M}}$), $[\mathrm{mol/m^3}]$}
\entry{$c_{s,p}^{\alpha},c_{s,p}^{\beta}$}{Positive particle solid phase concentration for $\alpha$ and $\beta$ phases,  $[\mathrm{mol/m^3}]$}
\entry{$k_{i}$}{Reaction rate ($i \in\hat{\mathcal{M}}$), $[\mathrm{m^{2.5}/(mol^{0.5}\cdot s)}]$}
\entry{$J_{i}$}{Pore wall flux ($i \in\hat{\mathcal{M}}$), $[\mathrm{mol/(m^3\cdot s)}]$}
\entry{$I$}{Applied current, $[\mathrm{A}]$}
\entry{$Q_{C/12}^k$}{Charge and discharge capacity at C/12 ($k\in\mathcal{K}$), $[\mathrm{Ah}]$}
\entry{$i_{0,i}$}{Exchange current ($i \in\hat{\mathcal{M}}$), $[\mathrm{A/m^2}]$}
\entry{$\kappa$}{Electrolyte phase conductivity ($i \in\mathcal{M}$), $[\mathrm{S/m}]$}
\entry{$\kappa_{eff,i}$}{Effective electrolyte phase conductivity ($i \in\mathcal{M}$), $[\mathrm{S/m}]$}
\entry{$R_{l}$}{Lumped contact resistance, $[\Omega]$}
\entry{$R_{el}$}{Electrolyte resistance, $[\Omega]$}
\entry{$\Phi_{s,i}$}{Solid phase potential ($i \in\hat{\mathcal{M}}$), $[\mathrm{V}]$}
\entry{$\Delta\Phi_e$}{Diffusion overpotential, $[\mathrm{V}]$}
\entry{$\phi_{e}$}{Electrolyte phase potential, $[\mathrm{V}]$}
\entry{$U_n$}{Negative electrode open circuit potential, $[\mathrm{V}]$}
\entry{$U_p^{ch},U_p^{dis}$}{Charge and discharge positive electrode open circuit potentials,  $[\mathrm{V}]$}
\entry{$V$}{Cell voltage, $[\mathrm{V}]$}
\entry{$\eta_i$}{Overpotential ($i \in\hat{\mathcal{M}}$), $[\mathrm{V}]$}
\entry{$T$}{Battery cell temperature, $[\mathrm{K}]$}
\entry{$R$}{Universal gas constant, $[\mathrm{J/(mol\cdot K)}]$}
\entry{$F$}{Faraday constant, $[\mathrm{C/mol}]$}

\section*{Notation}
\entry{$\hspace{0.2em}\widehat{}$}{Optimal solver setting $\mathcal{C}$} 
\entry{$\star$}{Optimal parameter} 
\entry{$y$}{Scalar variable, parameter} 
\entry{$\underline{y},\overline{y}$}{Lower and upper bound for $y$} 
\entry{$\mathbf{y}$}{Vector of variables $y$} 
\entry{$\mathbf{Y}$}{Matrix} 
\entry{$\iota$}{Index indicating the $\iota$-th model parameter vector}
\entry{$i$}{Index indicating the media: $p$, $n$, or $s$}
\entry{$j$}{Index indicating time}
\entry{$k$}{Index indicating charge and discharge scenarios}
\entry{$l$}{Index indicating a discretization point}
\entry{$e$}{Error estimate} 
\entry{$\mathrm{reltol},\mathrm{abstol}$}{Relative and absolute tolerance}
\entry{$\mathcal{C}$}{Solver setting}
\entry{$\Theta$}{Model parameter vector}
\entry{$\mathcal{K}$}{Set defined as $\{\mathrm{charge},\mathrm{discharge}\}$}
\entry{$\mathcal{M}$}{Set defined as $\{p,s,n\}$}
\entry{$\hat{\mathcal{M}}$}{Set defined as $\{p,n\}$}
\entry{$\mathrm{g}(I)$}{Concentration at the boundary}
\entry{$\mathrm{ic}_k$}{Core initial condition}
\entry{$\mathrm{sign}$}{Sign function}
\entry{$\alpha$}{Positive particle $\alpha$-phase}
\entry{$\beta$}{Positive particle $\beta$-phase}
\entry{$n$}{Negative particle}
\entry{$p$}{Positive particle}
\entry{$s$}{Separator}
\entry{$avg$}{Average along $x$}
\entry{$bulk$}{Particle bulk}
\entry{$ch,dis$}{Charge and discharge}
\entry{$eff$}{Effective}
\entry{$exp$}{Experimental}
\entry{$filler$}{Filler property}
\entry{$max$}{Maximum}
\entry{$surf$}{Particle surface}
\entry{$tot$}{Total}

\begin{table}[t]
	\caption{Governing equations of the core-shell ESPM model.}\label{table:ESPM_table_1}
	\centering
	\resizebox{0.95\textwidth}{!}{	
	\scriptsize{				
		\begin{tabular}{ll}
			\toprule\toprule \\ 
			\textbf{Negative electrode} $(n)$ & \hspace{1em}\textbf{Boundary conditions}  \\ 
			\parbox{9cm}{
			    \begin{equation*}
			    \begin{split}
			    & \varepsilon_n\frac{\partial c}{\partial t} = \frac{\partial}{\partial x}\left(D_n^{eff}(c,T)\frac{\partial c}{\partial x}\right) + (1-t_+)J_n,\ \ J_n = \frac{I}{A_{cell}FL_n}\hspace{10em}\\
			    & 
			    \end{split}
				\end{equation*}} & 
			\parbox{9cm}{
				\begin{equation} \label{eq:eq_ne1}
			    \begin{split}
			    & \frac{\partial c}{\partial x}\bigg\vert_{x=0}=0 \\
			    & D_n^{eff}(c,T)\frac{\partial c}{\partial x}\bigg\vert_{x = L_n^-} = D_s^{eff}(c,T)\frac{\partial c}{\partial x}\bigg\vert_{x = L_n^+}\hspace{7em}
			    \end{split}
				\end{equation}}\\[-1.5em]
		    \parbox{9cm}{
			    \begin{equation*}
			    \begin{split}
			    &\kappa_{eff,n}(c)\frac{\partial}{\partial x}\left(\frac{\partial \phi_e}{\partial x}\right) - \frac{2RT\kappa_{eff,n}(c)v(c,T)}{F}\frac{\partial^2\ln(c)}{\partial x^2}+FJ_n=0\hspace{1.5em}\\ 
			    &
			    \end{split}
				\end{equation*}} & 
			\parbox{9cm}{
				\begin{equation} \label{eq:eq_ne2}
			    \begin{split}
			    & \frac{\partial\phi_e}{\partial x}\bigg\vert_{x=0}=0\\
			    & \kappa_{eff,n}(c) \frac{\partial\phi_e}{\partial x}\bigg\vert_{x=L_n^-} = \kappa_{eff,s}(c) \frac{\partial\phi_e}{\partial x}\bigg\vert_{x=L_n^+}\hspace{7em}
			    \end{split}\raisetag{2.25\baselineskip}
				\end{equation}}\\[-1.5em]
		    \parbox{9cm}{
			    \begin{equation*}
			    \begin{split}
			    &\frac{\partial c_{s,n}}{\partial t} = D_{s,n}\frac{\partial^2c_{s,n}}{\partial r^2}+\frac{2D_{s,n}}{r}\frac{\partial c_{s,n}}{\partial r}\hspace{16em}\\
			    &
			    \end{split}
				\end{equation*}} & 
			\parbox{9cm}{
				\begin{equation} \label{eq:eq_ne3}
			    \begin{split}
			    & \frac{\partial c_{s,n}}{\partial r}\bigg\vert_{r=0}=0\\
			    & \frac{\partial c_{s,n}}{\partial r}\bigg\vert_{r = R_n} = \frac{-I}{D_{s,n}a_{n}A_{cell}FL_n}\hspace{12em}
			    \end{split}
				\end{equation}}\\ 
				
			\textbf{Separator} $(s)$ &  \\
			\parbox{9cm}{
			    \begin{equation*}
			    \begin{split}
			    & \varepsilon_s\frac{\partial c}{\partial t} = \frac{\partial}{\partial x}\left(D_s^{eff}(c,T)\frac{\partial c}{\partial x}\right)\hspace{18em}\\
			    &
			    \end{split}
				\end{equation*}} & 
			\parbox{9cm}{
				\begin{equation} \label{eq:eq_s1}
			    \begin{split}
			    & c\big\vert_{x=L_n^-}=c\big\vert_{x=L_n^+},\ c\big\vert_{x=L_n+L_s^-}=c\big\vert_{x=L_n+L_s^+} \hspace{6em}\\
			    & 
			    \end{split}
				\end{equation}}\\[-2.5em]
		    \parbox{9cm}{
			    \begin{equation*}
			    \begin{split}
			    &\kappa_{eff,s}(c)\frac{\partial}{\partial x}\left(\frac{\partial \phi_e}{\partial x}\right) - \frac{2RT\kappa_{eff,s}(c)v(c,T)}{F}\frac{\partial^2\ln(c)}{\partial x^2}=0\hspace{6em}\\
			    &
			    \end{split}
				\end{equation*}} & 
			\parbox{9cm}{
				\begin{equation} \label{eq:eq_s2}
			    \begin{split}
			    \phi_e\big\vert_{x=L_n^-}=\phi_e\big\vert_{x=L_n^+},\ \phi_e\big\vert_{x=L_n+L_s^-}=\phi_e\big\vert_{x=L_n+L_s^+}\hspace{4em}&\\
			    &
			    \end{split}\raisetag{2.25\baselineskip}
				\end{equation}}\\ 
				
           \textbf{Positive electrode} $(p)$ &  \\
			\parbox{9cm}{
			    \begin{equation*}
			    \begin{split}
			    & \varepsilon_p\frac{\partial c}{\partial t} = \frac{\partial}{\partial x}\left(D_p^{eff}(c,T)\frac{\partial c}{\partial x}\right) + (1-t_+)J_p,\ \ J_p = -\frac{I}{A_{cell}FL_p}\hspace{14em}\\
			    &
			    \end{split}
				\end{equation*}} & 
			\parbox{9cm}{
				\begin{equation} \label{eq:eq_pe1}
			    \begin{split}
			    & \frac{\partial c}{\partial x}\bigg\vert_{x=L_n+L_s+L_p}=0 \\
			    & D_s^{eff}(c,T)\frac{\partial c}{\partial x}\bigg\vert_{x =  L_n+L_s^-} = D_p^{eff}(c,T)\frac{\partial c}{\partial x}\bigg\vert_{x = L_n+L_s^+}\hspace{4em}
			    \end{split}
				\end{equation}}\\[-1.5em]
		    \parbox{9cm}{
			    \begin{equation*}
			    \begin{split}
			    &\kappa_{eff,p}(c)\frac{\partial}{\partial x}\left(\frac{\partial \phi_e}{\partial x}\right) - \frac{2RT\kappa_{eff,p}(c)v(c,T)}{F}\frac{\partial^2\ln(c)}{\partial x^2}+FJ_p=0\hspace{3em}\\
			    &
			    \end{split}
				\end{equation*}} & 
			\parbox{9cm}{
				\begin{equation} \label{eq:eq_pe2}
			    \begin{split}
			    & \frac{\partial\phi_e}{\partial x}\bigg\vert_{x=L_n+L_s+L_p}=0\\
			    & \kappa_{eff,s}(c) \frac{\partial\phi_e}{\partial x}\bigg\vert_{x=L_n+L_s^-} = \kappa_{eff,p}(c) \frac{\partial\phi_e}{\partial x}\bigg\vert_{x= L_n+L_s^+}\hspace{4.5em}
			    \end{split}
				\end{equation}}\\[-1.5em]
		    \parbox{9cm}{
			    \begin{equation*}
			    \begin{split}
			    &\frac{\partial c_{s,p}}{\partial t} = D_{s,p}\frac{\partial^2c_{s,p}}{\partial r^2}+\frac{2D_{s,p}}{r}\frac{\partial c_{s,p}}{\partial r}\hspace{16em}\\
			    \end{split}
				\end{equation*}} & 
			\parbox{9cm}{
				\begin{equation} \label{eq:eq_pe3}
			    \begin{split}
			    & \frac{\partial c_{s,p}}{\partial r}\bigg\vert_{r=0}=0\\
			    & \frac{\partial c_{s,p}}{\partial r}\bigg\vert_{r = R_p} = \frac{I}{D_{s,p}a_{p}A_{cell}FL_p}\hspace{13em}
			    \end{split}
				\end{equation}}\\[-1.5em]
			\parbox{9cm}{
			    \begin{equation*}
			    \begin{split}
			    & \text{One-phase:}\quad \frac{d r_p}{d t} = 0\hspace{30em}\\
			    \end{split}
				\end{equation*}} & 
			\parbox{9cm}{
				\begin{equation} \label{eq:eq_pe5}
			    \begin{split}
			    &r_p\big|_{t=0} = 0\hspace{22em}\\
			    \end{split}
				\end{equation}}\\[-1.5em]
				\parbox{9cm}{
			    \begin{equation*}
			    \begin{split}
			    &\text{Two-phase:}\quad \mathrm{sign}(I)(c_{s,p}^\alpha-c_{s,p}^\beta)\frac{d r_p}{d t} = D_{s,p}\frac{\partial c_{s,p}}{\partial r}\bigg|_{r = r_p}\hspace{6em}\\
			    \end{split}
				\end{equation*}} & 
			\parbox{9cm}{
				\begin{equation} \label{eq:eq_pe4}
			    \begin{split}
			    &r_p\big|_{t=\bar{t}} = R_p-\epsilon, \ \ c_{s,p}\big|_{r=r_p} = \mathrm{g}(I), \ \ c_{s,p}\big|_{t=\bar{t}\wedge r\in[0,R_p]} = \mathrm{ic}_k\hspace{1em}\\
			    \end{split}
				\end{equation}}\\
			\bottomrule \bottomrule
		\end{tabular}}}
\end{table}

\begin{table}[!htb]
	\caption{Additional equations for the core-shell ESPM model (part 1).}\label{table:ESPM_table_2}
      \centering
	\resizebox{0.7\columnwidth}{!}{					
		\begin{tabular}{l}
			\toprule\toprule \\ [-1mm]
		     \multicolumn{1}{l}{\textbf{Current convention}}  \\ [-2mm]
			\parbox{12cm}{
			   \begin{flalign} \label{eq:currconv_1}
				& \quad\begin{cases}
				I > 0, \ \ \mathrm{discharge}\\
				I = 0, \\
				I < 0, \ \ \mathrm{charge}\\
				\end{cases} &
				\end{flalign}}\\[0mm]
				
			 \multicolumn{1}{l}{\textbf{Concentration at the moving boundary}}  \\ [-2mm]
		    	\parbox{12cm}{
			   \begin{flalign} \label{eq:currconv_2}
				&  \quad\mathrm{g}(I) = \begin{cases}
				c_{s,p}^\beta = \theta_p^\beta\cdot c_{s,p}^{max}, \ \ \text{if}\ I > 0\\
	             c_{s,p}^\alpha  = \theta_p^\alpha\cdot c_{s,p}^{max}, \ \text{if}\ I < 0\\
				0,\ \ \text{otherwise}\\
				\end{cases}  &
				\end{flalign}}\\[0mm]
				
			\multicolumn{1}{l}{\textbf{Core-shell initial condition}}  \\ [-2mm]
			\parbox{12cm}{
			    \begin{flalign} \label{eq:currconv_3}
				& \quad\mathrm{ic}_k = \begin{cases}
	              c_{s,p}^\alpha,\quad k=\mathrm{discharge}\\
	              c_{s,p}^\beta,\quad k=\mathrm{charge}\\
				\end{cases} &
				\end{flalign}}\\[0mm]
				
			\multicolumn{1}{l}{\textbf{Diffusivity and conductivity}}  \\ [-2mm]
			\parbox{12cm}{
			    \begin{flalign} \label{eq:diff_1}
				& \quad D_i^{eff}(c,T) = D(c,T)\cdot \varepsilon_i^{brugg},\ \ i \in\mathcal{M} &
				\end{flalign}
				\begin{flalign*}
				& \quad\quad\rightarrow D(c,T) = 0.0001\cdot 10^{\left(-4.51-\frac{59.22}{T-(206.25+10c/1000)}\right)c/1000} &
				\end{flalign*}} \\[-1mm]
			\parbox{12cm}{
			    \begin{flalign} \label{eq:diff_2}
				& \quad \kappa_{eff,i}(c) = \kappa(c)\cdot \varepsilon_i^{brugg}, \ \ i \in\mathcal{M} &
				\end{flalign}
				\begin{flalign*}
				& \quad\quad\rightarrow \kappa(c) = \left(\frac{c^{avg}/1000}{1.05}\right)^{0.68} 
				 \mathrm{exp}[-0.1(c^{avg}/1000-1.05)^2+\\ &\hspace{5.6em}- 0.56\left(c^{avg}/1000-1.05\right)] &
				\end{flalign*}} \\[0mm]
				
		     \multicolumn{1}{l}{\textbf{Active area}}  \\ [-2mm]
			\parbox{12cm}{
			    \begin{flalign} \label{eq:act_area}
				& \quad a_i=\frac{3}{R_i}\nu_i,\ \ i \in\hat{\mathcal{M}} &
				\end{flalign}} \\[0mm]
				
			\multicolumn{1}{l}{\textbf{Porosity}}  \\ [-2mm]
			\parbox{12cm}{
			    \begin{flalign} \label{eq:poro_1}
				& \quad \varepsilon_{i} = 1-\nu_i-\nu_{i,filler},\ \ i \in\hat{\mathcal{M}} &
				\end{flalign}} \\[8mm]
				
			\bottomrule \bottomrule
		\end{tabular}}
\end{table}

\begin{table}[!htb]
	\caption{Additional equations for the core-shell ESPM model (part 2).}\label{table:ESPM_table_2b}
      \centering
	\resizebox{0.7\columnwidth}{!}{					
		\begin{tabular}{l}
			\toprule\toprule \\ [-1mm]
			\multicolumn{1}{l}{\textbf{Cell voltage}}  \\ [-2mm]
			\parbox{12cm}{
			    \begin{flalign} \label{eq:cell_volt_3}
				& \quad  \Phi_{s,i} =U_i(\theta_i^{surf}) + \eta_i,\ \ i \in\hat{\mathcal{M}} &&
				\end{flalign}} \\[-2mm]
		     \parbox{12cm}{
			    \begin{flalign} \label{eq:cell_volt_4}
				& \quad  \Delta\Phi_e = \frac{2RTv(c,T)}{F}\ln\left(\frac{c(L)}{c(0)}\right),\ \mathrm{with}\  L = L_n+L_s+L_p& 
				\end{flalign} \vspace{-2.2em}
				\begin{flalign*}
		         & \quad\quad\rightarrow v(c,T) = 0.601 - 0.24(c^{avg}/1000)^{1/2} + \\ &\quad\hspace{4.5em} + 0.982\left[1-0.0052(T-293)\right](c^{avg}/1000)^{3/2}\ \text{\cite{tanim2015temperature}} &
				\end{flalign*}} \\[-2mm]
			\parbox{12cm}{
			    \begin{flalign} \label{eq:cell_volt_1}
				& \quad V = \Phi_{s,p}-\Phi_{s,n}+\Delta\Phi_{e} - I(R_l+R_{el}) &
				\end{flalign}} \\[-2mm]
			\parbox{12cm}{
			    \begin{flalign} \label{eq:cell_volt_2}
				& \quad R_{el} = \frac{1}{2A_{cell}} \left(\frac{L_n}{\kappa_{eff,n}(c)}+\frac{2L_s}{\kappa_{eff,s}(c)}+\frac{L_p}{\kappa_{eff,p}(c)}\right) &
				\end{flalign}} \\[-2mm]
	         \parbox{12cm}{
			    \begin{flalign} \label{eq:cell_volt_5}
				& \begin{cases}
				\begin{split} U_p^{dis} &= 3.382 -0.2955 \exp{\left[-44.99(1-\theta_{p}^{surf})^{0.8707}\right]} +\\
    &+ 10^{-20.71} \exp{\left[14.17 (1-\theta_{p}^{surf})^{8.128}\right]} +\\[1mm]
   &+ 10^{-40.82} \exp{\left[100(1-\theta_{p}^{surf})^{1.213}\right]}\end{split}, \quad \mathrm{discharge}\\
				\begin{split}     U_p^{ch} &= 3.442 -0.1774 \exp{\left[-127.7(1-\theta_{p}^{surf})^{0.7921}\right]} +\\
    &+ 10^{-2.123}\exp{\left[16.56 (1-\theta_{p}^{surf})^{24.08}\right]}+ \\[1mm]
    &+ 10^{-10.29}\exp{\left[99.91(1-\theta_{p}^{surf})^{22.17}\right]}\end{split},\quad \mathrm{charge} \end{cases}&
				\end{flalign}} \\[-5mm]
				\parbox{12cm}{
			    \begin{flalign} \label{eq:cell_volt_6}
                 & \quad\theta_{n}^{surf}  = c_{s,n}/c_{s,n}^{max},\quad  \theta_{p}^{surf}  = c_{s,p}/c_{s,p}^{max} & 
				\end{flalign}} \\[-1mm]				
				
			\multicolumn{1}{l}{\textbf{Electrochemical overpotential}}  \\ [-2mm]
			\parbox{12cm}{
			    \begin{flalign} \label{eq:overp_1}
				& \quad \eta_i =  \frac{RT}{0.5F}\sinh^{-1}\left(\frac{I}{2A_{cell}\ a_{i}\ L_i\ i_{0,i}}\right),\ \ i \in\hat{\mathcal{M}} &
				\end{flalign}} \\[-2mm]
			\parbox{12cm}{
			    \begin{flalign} \label{eq:overp_2}
				& \quad i_{0,i} = k_iF\sqrt{c^{avg}c_{s,i}^{surf}\left(c_{s,i}^{max}-c_{s,i}^{surf}\right)},\ \ i \in\hat{\mathcal{M}} &
				\end{flalign}} \\[-1mm]				
		
		    \multicolumn{1}{l}{\textbf{State of charge}}  \\ [-2mm]
			\parbox{12cm}{
			    \begin{flalign} \label{eq:soc_1}
                  \begin{split}
                  &  SOC_n = \frac{\theta_n^{bulk}-\theta_{n,0\%}}{\theta_{n,100\%}-\theta_{n,0\%}},\ SOC_p = \frac{\theta_{p,0\%}-\theta_p^{bulk}}{\theta_{p,0\%}-\theta_{p,100\%}}\hspace{5em}\\ 
                  & \mathrm{Negative\ particle}: \quad \theta_{n}^{bulk}  = \frac{3}{c_{s,n}^{max}R_n^3}\int_{0}^{R_n}c_{s,n}r^2dr\\
                  & \mathrm{Positive\ particle}: \quad  \theta_{p}^{bulk}  = \frac{3}{c_{s,p}^{max}R_p^3}\int_{0}^{R_p}c_{s,p}r^2dr
                  \end{split}
				\end{flalign}} \\[-2mm]
			\bottomrule \bottomrule
		\end{tabular}}
\end{table}

\renewcommand{\arraystretch}{1}
\begin{table}[!tb]
	\caption{State-space representation for mass transport in the electrolyte and solid phases.}	
	\centering
	\label{tab:discradditional}		
	\resizebox{1\columnwidth}{!}{					
		\rev{\begin{tabular}{l}
			\toprule\toprule \\ [-1mm]
		    \multicolumn{1}{l}{\textbf{Mass transport in the electrolyte phase} \cite{weaver2020novel}}  \\ [-2mm]
			\parbox{20cm}{
			    \begin{flalign} \label{eq:ssadd1}
				\begin{split}
				& \dot{\mathbf{c}} = \mathbf{A}_e\mathbf{c} + \frac{(1-t_+)}{A_{cell}F} \mathbf{B}_eI,\ \ \mathbf{c} = \begin{bmatrix} \mathbf{c}_n\\ \mathbf{c}_s \\ \mathbf{c}_p\end{bmatrix}_{N_{x,tot}\times 1}\in\mathbb{R}^{N_{x,tot}\times 1}, \ \text{with} \quad N_{x,tot} = \sum\limits_{i\in\mathcal{M}}N_{x,i} \\[2mm]
				& \mathbf{A}_{e} = \begin{bmatrix}
					-{D_{n,1/2}^{eff}} & {D_{n,1/2}^{eff}} & 0 & 0 & \dots & 0  \\
					{D_{n,1/2}^{eff}}& -(D_{n,1/2}^{eff}+D_{n,3/2}^{eff}) & {D_{n,3/2}^{eff}} & 0 & \dots & 0\\
					0 & {D_{n,3/2}^{eff}}& -(D_{n,3/2}^{eff}+D_{n,5/2}^{eff}) & {D_{n,5/2}^{eff}}  & \dots & 0\\
					0 & 0 & {D_{n,5/2}^{eff}}& -(D_{n,5/2}^{eff}+D_{n,7/2}^{eff})  & \dots & 0\\
					\vdots & \vdots & \vdots & \vdots & \ddots & \vdots \\
					0 & 0 & 0 & 0 & \dots &  -{D_{p,N_{x,tot}-3/2}^{eff}}\\
					\end{bmatrix}_{N_{x,tot}\times N_{x,tot}}\\
					& \mathbf{B}_{e} = 
					\begin{bmatrix}
					\frac{(1)_{N_{x,i}\times 1}}{L_n}\\[1mm]
					\frac{(0)_{N_{x,i}\times 1}}{L_s}\\[1mm]
					\frac{(-1)_{N_{x,i}\times 1}}{L_p}\\[1mm]
					\end{bmatrix}_{N_{x,tot}\times 1}\\[2mm]
			     & \Delta_{x_i}=\frac{L_i}{N_{x,i}}, \ \overline{\Delta}=\frac{\Delta_{x_i,l}}{\Delta_{x_i,l-1}+\Delta_{x_i,l}},  \ D_{i,l\pm 1/2}^{eff}=\frac{D_{i,l}^{eff}(c,T)D_{i,l\pm 1}^{eff}(c,T)}{\overline{\Delta}D_{i,l}^{eff}(c,T)+(1-\overline{\Delta})D_{i,l\pm 1}^{eff}(c,T)},\ \mathrm{with}\ i\in\mathcal{M}\ \mathrm{and}\ l\in[0,N_{x,tot}-1]\\
				\end{split}
				\end{flalign}} \\[7mm]
				
		    \multicolumn{1}{l}{\textbf{Mass transport in the solid phase - negative electrode} \cite{weaver2020novel}}  \\ [-2mm]
			\parbox{20cm}{
			    \begin{flalign} \label{eq:ssadd2}
				\begin{split}
				& \dot{\mathbf{c}}_{s,n} = \frac{D_{s,n}}{(R_n/(N_{r,n}-1))^2}\mathbf{A}_{s,n}\mathbf{c}_{s,n} +  \frac{-1}{A_{cell}L_nFa_{n}(R_n/(N_{r,n}-1))}\mathbf{B}_{s,n}I,\ \ \mathbf{c}_{s,n}\in\mathbb{R}^{(N_{r,n}-1)\times 1}\\[2mm]
				&\mathbf{A}_{s,n} = 
					\begin{bmatrix}
					-2 & 2 & 0 & 0 & \dots & 0 \\
					1/2 & -2 & 3/2 & 0 & \dots & 0 \\
					0 & 2/3 & -2 & 4/3 & \dots & 0 \\
					0 & 0 & 3/4 & -2 & \dots & 0 \\
					\vdots & \vdots & \vdots & \vdots & \ddots & \vdots \\
					0 & 0 & 0 & 0 & \dots & -2\\
					\end{bmatrix}_{(N_{r,n}-1)\times (N_{r,n}-1)}, \quad \mathbf{B}_{s,n} = 
					\begin{bmatrix}
					0\\
					0\\ 
					0\\
					0\\
					\vdots\\
					\left(2+\frac{2}{N_{r,n}-1} \right)\\
					\end{bmatrix}_{(N_{r,n}-1)\times 1}\hspace{15.5em}
				\end{split}
				\end{flalign}} \\[7mm]
				
		     \multicolumn{1}{l}{\textbf{Mass transport in the solid phase - positive electrode} (Section \ref{sec:ssposel})}  \\ [-2mm]
			\parbox{20cm}{
			    \begin{flalign} \label{eq:ssadd3}
				\begin{split}
				&\dot{\mathbf{x}} = \eta_1\mathbf{A}_{s,p1}\mathbf{x} + \eta_2\mathbf{A}_{s,p2}\mathbf{x} + \eta_3\mathbf{B}_{s,p}I+\eta_1\mathbf{G}_{s,p}, \ \ \mathbf{x}=\begin{bmatrix} r_p\\ \mathbf{c}_{s,p}\end{bmatrix}\in\mathbb{R}^{N_{r,p}\times 1}\\[2mm]
				&\mathbf{A}_{s,p1} = 
\begin{bmatrix}
0 & \eta_4/\eta_1  & 0 & 0 & 0 & \dots & 0  \\
0 & -2 & 1 & 0 & 0 & \dots & 0  \\
0 &1 & -2 & 1 & 0 & \dots & 0 \\
0 &0 & 1 & -2 & 1 & \dots & 0 \\
0 &0 & 0 & 1 & -2 & \dots & 0 \\
\vdots & \vdots & \vdots & \vdots & \vdots & \ddots & \vdots\\
0& 0 & 0 & 0 & 0 & \dots &  -1\\
\end{bmatrix}_{N_{r,p}\times N_{r,p}}, \quad \mathbf{A}_{s,p2} = 
\begin{bmatrix}
0 & 0 & 0 & 0 & 0 & \dots & 0 \\
0 & -1 & 1 & 0 & 0 & \dots & 0 \\
0 & 0 & -1 & 1 & 0 & \dots & 0 \\
0 &0 & 0 & -1 & 1 & \dots & 0 \\
0 &0 & 0 &  0 & -1 & \dots & 0 \\
\vdots & \vdots & \vdots & \vdots & \vdots & \ddots & \vdots \\
0 &0 & 0 & 0 & 0 & \dots & 0 \\
\end{bmatrix}_{N_{r,p}\times N_{r,p}}\hspace{5em}\\[2mm]
				& \mathbf{B}_{s,p} = 
\begin{bmatrix}
0\\
0\\
0\\
\vdots\\
0\\
1\\
\end{bmatrix}_{N_{r,p}\times 1}, \quad \mathbf{G}_{s,p} = 
\begin{bmatrix}
-\eta_4/\eta_1\mathrm{g}(I)\\
\mathrm{g}(I)\\
0\\ 
\vdots\\
0\\
0\\
\end{bmatrix}_{N_{r,p}\times 1}
				\end{split}
				\end{flalign}} \\[-2mm]
			\bottomrule \bottomrule
		\end{tabular}}}
\end{table}

\begin{table}[t!]
	\centering
	\begin{center}
		\caption{Charge and discharge capacities at C/12,  C/10,  C/6,  and C/3.}
		\label{tab:dischcapacities}
		\rev{\begin{tabular}{cccc}
			\toprule
			\textbf{C-rate} & \textbf{Charge capacity} & \textbf{Discharge capacity}  & \textbf{Unit} \\ 
		     \midrule
			C/12 &  49.935 ($Q_{C/12}^\text{charge}$) & 49.999 ($Q_{C/12}^\text{discharge}$) &  $[\mathrm{Ah}]$\\ 
			C/10 & 49.866 & 49.952  &$[\mathrm{Ah}]$\\ 
			C/6 &49.711 &  49.775 &  $[\mathrm{Ah}]$\\ 
			C/3 &48.629 &  49.109 &  $[\mathrm{Ah}]$\\ 
		     \bottomrule
		\end{tabular}}
	\end{center}
\end{table}

\begin{table}[!tb]
	\caption{Identified core-shell ESPM parameters at different C-rate. Grey values are parameters not identified at the particular C-rate.  $k_n$ and $k_p$ are identified for C/6 and C/3, only.}	
	\centering
	\label{tab:id_results_all}
	\resizebox{\textwidth}{!}{	
	\begin{tabular}{l p{5em} p{5em} |c|cccc|c}
	\toprule
	\textbf{Parameter} & \textbf{Lower bound $\underline{y}$} &\textbf{Upper bound $\overline{y}$} & \rev{\begin{tabular}{c}\textbf{Initial}\\ \textbf{guess $\boldsymbol{\Theta}_1$}\vspace{-0.9em}\end{tabular}} & \rev{$\Theta_{C/12}$} &  \rev{$\Theta_{C/10}$}  &  \rev{$\Theta_{C/6}$}   &  \rev{$\Theta_{C/3}$}  & \textbf{Unit} \\
	\midrule
	$R_{n}$ & $1{\times}10^{-6}$ & $2{\times}10^{-5}$  & $9.3{\times}10^{-6}$ &  $\mathbf{1.03{\times}10^{-6}}$ & \textcolor{mygrey}{$1.03{\times}10^{-6}$} & \textcolor{mygrey}{$1.03{\times}10^{-6}$} & \textcolor{mygrey}{$1.03{\times}10^{-6}$} & $[\mathrm{m}]$ \\
	$R_{p}$ & $1{\times}10^{-8}$ & $1{\times}10^{-5}$  & $7.4{\times}10^{-7}$ & $\mathbf{4.32{\times}10^{-8}}$ & \textcolor{mygrey}{${4.32{\times}10^{-8}}$} &  \textcolor{mygrey}{${4.32{\times}10^{-8}}$} &  \textcolor{mygrey}{${4.32{\times}10^{-8}}$} & $[\mathrm{m}]$ \\
	$A_{cell}$ & $1.41$ & $1.73$  & $1.57$ & \textbf{1.491} & \textcolor{mygrey}{ 1.491} &  \textcolor{mygrey}{ 1.491}  &  \textcolor{mygrey}{ 1.491}  & $[\mathrm{m^2}]$ \\		
	$D_{s,n}$ & $1{\times}10^{-15}$ & $1{\times}10^{-10}$  &  $1{\times}10^{-13}$ & $\mathbf{6.93{\times}10^{-12}}$  & $\mathbf{5.13{\times}10^{-11}}$ & $\mathbf{1{\times}10^{-10}}$ & $\mathbf{1{\times}10^{-10}}$ & $[\mathrm{m^2/s}]$ \\
	$D_{s,p}$ & $1{\times}10^{-18}$ & $1{\times}10^{-11}$  & $1{\times}10^{-15}$& $\mathbf{3.11{\times}10^{-17}}$  & $\mathbf{3.12{\times}10^{-17}}$ & $\mathbf{3.12{\times}10^{-17}}$ & $\mathbf{4.65{\times}10^{-17}}$ & $[\mathrm{m^2/s}]$ \\	
	{$\theta_{n ,100\%}$} &  $\mathrm{0.7}$ & $\mathrm{0.95}$  &  $0.930$ & \textbf{0.835} & \textcolor{mygrey}{0.835} & \textcolor{mygrey}{0.835} & \textcolor{mygrey}{0.835} & $[\mathrm{-}]$ \\
	{$\theta_{n ,0\%}$} &  $1{\times}10^{-4}$ & $\mathrm{0.2}$ & $0.070$ & \textbf{0.0095} & \textcolor{mygrey}{0.0095} & \textcolor{mygrey}{0.0095} & \textcolor{mygrey}{0.0095} &  $[\mathrm{-}]$ \\
	{$\theta_{p ,100\%}$} &  $\mathrm{0.05}$ & $\mathrm{0.15}$ & $0.050$ & \textbf{0.0696} & \textcolor{mygrey}{0.0696}  & \textcolor{mygrey}{0.0696} & \textcolor{mygrey}{0.0696} & $[\mathrm{-}]$ \\
	{$\theta_{p ,0\%}$} &  $\mathrm{0.8}$ & $\mathrm{1}$ & $0.880$  & \textbf{0.8821} & \textcolor{mygrey}{0.8821} & \textcolor{mygrey}{0.8821} & \textcolor{mygrey}{0.8821} & $[\mathrm{-}]$ \\			
	{$\theta_p^{\alpha}$} &  $\mathrm{0.1}$ & $\mathrm{0.2}$  &$0.150$  & \textbf{0.198} & \textcolor{mygrey}{0.198} & \textcolor{mygrey}{0.198} & \textcolor{mygrey}{0.198} & $[\mathrm{-}]$ \\
	{$\theta_p^{\beta}$} &  $\mathrm{0.8}$ & $\mathrm{0.9}$ & $0.850$ & \textbf{0.8} & \textcolor{mygrey}{0.8} & \textcolor{mygrey}{0.8} & \textcolor{mygrey}{0.8} & $[\mathrm{-}]$ \\	
	$R_{l}$ & $1{\times}10^{-3}$ & $\mathrm{0.01}$  & $0.005$ & \textbf{0.001} & \textbf{0.001} & \textbf{0.001} &  \textbf{0.002}& $[\Omega]$ \\
	$k_n$ & $1{\times}10^{-14}$ & $1{\times}10^{-8}$ &  - & - & - & $\mathbf{7.70{\times}10^{-13}}$ & $\mathbf{3.83{\times}10^{-12}}$ & $[\mathrm{m^{2.5}/(mol^{0.5}s)}]$ \\
	$k_p$ & $1{\times}10^{-14}$ & $1{\times}10^{-8}$ &  - & - & - & $\mathbf{4.60{\times}10^{-14}}$ & $\mathbf{4.41{\times}10^{-10}}$ & $[\mathrm{m^{2.5}/(mol^{0.5}s)}]$ \\
	\bottomrule
	\end{tabular}}
\end{table}

\begin{table}[!tb]
	\caption{Values of the cost function $J$ at different C-rate for charge and discharge (identification).}	
	\centering
	\label{tab:id_costs}
	\begin{tabular}{c|cccc|cccc|c}
	\toprule
     & \multicolumn{4}{c|}{\textcolor{myred}{\textbf{Charge}}} & \multicolumn{4}{c|}{\textcolor{myred}{\textbf{Discharge}}} & \textbf{Unit} \\
     & \textbf{C/12} &  \textbf{C/10} &  \textbf{C/6}  &  \textbf{C/3} & \textbf{C/12} &  \textbf{C/10} &  \textbf{C/6}  &  \textbf{C/3} & \\
	\midrule
	$J_V$ & 0.0021 & 0.0022 & 0.0042 & 0.0097 &  0.0031& 0.0035& 0.0054& 0.0215& [-] \\
	$J_{SOC_n}$ & 0.0007 & 0.0005 & 0.0017 & 0.0039 & 0.0007 & 0.0006& 0.0006& 0.0023& [-]\\
	$J_{SOC_p}$ & 0.0034 & 0.0038 & 0.0025 & 0.0022 & 0.0010 & 0.0020& 0.0014& 0.0047& [-]\\
	$\boldsymbol{J}$ & \textbf{0.0062} & \textbf{0.0065}  & \textbf{0.0084} &\textbf{0.0158}  & \textbf{0.0048}  & \textbf{0.0061} & \textbf{0.0074} & \textbf{0.0285} & [-] \\
	\bottomrule
	\end{tabular}
\end{table}

\begin{table}[!tb]
	\caption{Values of the cost function $J$ at different C-rate for charge and discharge (verification).}	
	\centering
	\label{tab:val_costs}
	\begin{tabular}{c|cccc|cccc|c}
	\toprule
     & \multicolumn{4}{c|}{\textcolor{myred}{\textbf{Charge}}} & \multicolumn{4}{c|}{\textcolor{myred}{\textbf{Discharge}}} & \textbf{Unit} \\
     & \textbf{C/12} &  \textbf{C/10} &  \textbf{C/6}  &  \textbf{C/3} & \textbf{C/12} &  \textbf{C/10} &  \textbf{C/6}  &  \textbf{C/3} & \\
	\midrule
	$J_V$ & 0.0021& 0.0024& 0.0043& 0.0098&  0.0065& 0.0056 & 0.0105& 0.0294& [-] \\
	$J_{SOC_n}$ & 0.0029& 0.0027& 0.0040& 0.0060& 0.0020& 0.0016& 0.0021& 0.0041& [-]\\
	$J_{SOC_p}$ & 0.0025& 0.0025& 0.0029& 0.0051& 0.0031& 0.0023& 0.0023& 0.0052& [-]\\
	$\boldsymbol{J}$ & \textbf{0.0075} & \textbf{0.0076}  & \textbf{0.0112} &\textbf{0.0209}  & \textbf{0.0116}  & \textbf{0.0095} & \textbf{0.0149} & \textbf{0.0387} & [-] \\
	\bottomrule
	\end{tabular}
\end{table}

\renewcommand{\arraystretch}{1.3}
\begin{table}[!tb]
	\caption{Solver settings for the sensitivity analysis of numerical solutions,  \textit{Step 1.}}	
	\centering
	\label{tab:sens1}
	\begin{tabular}{lcc}
	\toprule
	\textbf{Parameters} & \textbf{Values tested}  & \textbf{Unit}\\
	\midrule
	$N_{r}$ & $\{5,10,20,30,40,50,60,70,80,90,100\}$ & [-] \\
	$dt$ & $\{1,10,20,30,40,50\}$ & [s]  \\
	$\mathrm{reltol}$ & $\{1{\times}10^{-9},1{\times}10^{-8},1{\times}10^{-7},1{\times}10^{-6},1{\times}10^{-5}\}$ & [-]  \\
	$\mathrm{abstol}$ & $\{ \mathrm{reltol}\times0.001, \mathrm{reltol}\times0.01,  \mathrm{reltol}\times0.1,\mathrm{reltol} \}$ &  \begin{tabular}{c}
	[$\mathrm{mol/m^3}$]\\[-1.5mm] (for $c_{s,i}$ and $c$)\\[-1mm] $\text{[}\mathrm{pm}\text{]}$ \\[-1.5mm]  (for $r_p$)\\
	\end{tabular}\\
	\bottomrule
	\end{tabular}
\end{table}

\renewcommand{\arraystretch}{1.3}
\begin{table}[!tb]
	\caption{Solver settings for the sensitivity analysis of numerical solutions,  \textit{Step 2.} Bold values are fixed from \textit{Step 1}.}	
	\centering
	\label{tab:sens2}
	\begin{tabular}{lcc}
	\toprule
	\textbf{Parameters} & \textbf{Values tested}  & \textbf{Unit}\\
	\midrule
	$N_{r}$ & $\{5,10,20,30,40,50,60,70,80,90,100\}$ & [-] \\
	$dt$ & $\mathbf{50}$ & [s]  \\
	$\mathrm{reltol}$ & $\{1{\times}10^{-9},1{\times}10^{-8},1{\times}10^{-7},1{\times}10^{-6},1{\times}10^{-5}\}$ & [-]  \\
	$\mathrm{abstol}$ & $\mathbf{{reltol}\times0.001}$ &  \begin{tabular}{c}
	[$\mathrm{mol/m^3}$]\\[-1.5mm] (for $c_{s,i}$ and $c$)\\[-1mm] $\text{[}\mathrm{pm}\text{]}$ \\[-1.5mm]  (for $r_p$)\\
	\end{tabular}\\
	\bottomrule
	\end{tabular}
\end{table}

\end{document}